\documentclass[aps,pra,twocolumn,groupedaddress]{revtex4}
\usepackage{graphicx}
\usepackage{amsmath}
\usepackage{amsfonts}
\usepackage{color}
\usepackage{bm}
\newcommand{\ket}[1]{\ensuremath{\left|{#1}\right\rangle}}
\newcommand{\bra}[1]{\ensuremath{\left\langle{#1}\right |}}

\newcommand{\oper}[1]{\mathbf{\mathsf{#1}}}

\newcommand{\sinc}{\ensuremath{{\mathrm{sinc}}}}

\newcommand{\vect}[1]{\ensuremath{\bm{\mathrm{#1}}}}

\begin{document}


\title{Continuous variable quantum computation with spatial degrees of freedom of photons}
\author{D. S. Tasca}
\email[]{tasca@if.ufrj.br} \affiliation{Instituto de F\'{\i}sica,
Universidade Federal do Rio de Janeiro, Caixa Postal 68528, Rio de
Janeiro, RJ 21941-972, Brazil}
\author{R. M. Gomes}
 \affiliation{Instituto de F\'{\i}sica, Universidade Federal do Rio
              de Janeiro, Caixa Postal 68528, Rio de Janeiro, RJ 21941-972,
              Brazil}
\author{F. Toscano}
 \affiliation{Instituto de F\'{\i}sica, Universidade Federal do Rio
              de Janeiro, Caixa Postal 68528, Rio de Janeiro, RJ 21941-972,
              Brazil}

\author{P. H. Souto Ribeiro}
\email[]{phsr@if.ufrj.br} \affiliation{Instituto de F\'{\i}sica,
Universidade Federal do Rio de Janeiro, Caixa Postal 68528, Rio de
Janeiro, RJ 21941-972, Brazil}
\author{S. P. Walborn}
\affiliation{Instituto de F\'{\i}sica, Universidade Federal do Rio
de Janeiro, Caixa Postal 68528, Rio de Janeiro, RJ 21941-972,
Brazil}
\date{\today}

\begin{abstract}
  We discuss the use of  the transverse spatial degrees of freedom of photons propagating in the paraxial approximation for continuous variable information processing. Given the wide variety of linear optical devices available, a diverse range of operations can be performed on the spatial degrees of freedom of single photons.   Here we show how to implement a set of continuous quantum logic gates which allow for universal quantum computation. In contrast with the usual quadratures of the electromagnetic field, the entire set of single photon gates for spatial degrees of freedom does not require optical nonlinearity and, in principle, can be performed with a single device: the \emph{spatial light modulator}. Nevertheless, nonlinear optical processes, such as four-wave mixing,  are needed in the implementation of two-photon gates. The efficiency of these gates is at present very low, however small scale investigations of continuous variable quantum computation are within the reach of current technology.  In this regard, we show how novel cluster states for one-way quantum computing can be produced using spontaneous parametric down conversion.     
\end{abstract}


\maketitle


\section{Introduction}
Quantum information is usually cast in terms of discrete two-level systems, which are used to encode qubits--the simplest form of quantum information.  In analogy with classical bits, a number of single and multiple-qubit logic gates can be used to perform quantum computation and other quantum information processing tasks \cite{chuang00}.   
There has been considerable interest in continuous variable (CV) degrees of freedom for quantum information processing \cite{lloyd99}.  Both the standard quantum computational model \cite{chuang00} and the one-way quantum computation  model \cite{raussendorf01} have been extended to continuous variable systems \cite{lloyd99,menicucci06,vanloock07,gu09,vanloock10}.  In both of these paradigms of CV quantum computation it is necessary to produce either non-gaussian states or implement non-gaussian operations \cite{lloyd99,menicucci06}.  In the standard model, Lloyd and Braunstein have shown that any single-mode gate that is cubic or higher order in the canonical variables is sufficient (in addition to gaussian operations) to implement universal quantum computation \cite{lloyd99}.  In terms of the one-way model, any single-mode non-gaussian measurement will suffice \cite{menicucci06}.  The typical example of quantum system with continuous variables are the quantized modes of the electromagnetic field, whose CV quadrature operators are analogous to the position and momentum operators of a  quantum harmonic oscillator \cite{braunstein05}. For quantum information tasks, each mode corresponds to the analog of a qubit, sometimes called ``qu-mode" \cite{vanloock10}.
  \par
The quantum CV formalism applies to any CV quantum system, and as such, there are a number of systems which can be used to explore quantum information processing.  In particular, the transverse spatial degrees of freedom of single photons present a rich playground for the investigation of quantum information in CV's.  Though the transverse modes of photons have been used to investigate quantum information for discrete variables \cite{vaziri02,vaziri03,oliveira05,abouraddy07,yarnall07}, to our knowledge they have been seldom explored in the context of CV quantum information processing.  In this case, the terminology ``qu-mode" might seem misleading, since we will exploit the spatial multi-mode structure of single-photon fields.  Nevertheless, there exists a direct analogy between operators which act in the single-mode multi-photon Hilbert  space of the electromagnetic field and
those that act in the single-photon multi-mode Hilbert space of the electromagnetic field \cite{abouraddy07}.    
A number of experimental and theoretical studies which exploit this analogy have been performed, including the experimental investigation and detection of two photon entanglement \cite{howell04,dangelo04,tasca08,tasca09,pires09c}, studies of quantum key distribution \cite{almeida05,walborn06,walborn08}, and simulation of stronger-than-quantum correlations \cite{tasca09b}. Moreover, there is the possibility of producing spatially non-gaussian entangled states by manipulating the pump beam \cite{gomes09b}, which have interesting properties \cite{walborn03a,walborn03b,caetano03,nogueira04a,gomes09a}. Furthermore, rotations and measurements in spatial parity space can be performed so as to implement a pseudo-spin 1/2 system \cite{abouraddy07}, and it was shown that the parity correlations of the spatial degrees of freedom of a pair of photons violate a Bell's inequality \cite{yarnall07}. There have also been a number of experimental investigations of spatial entanglement with the discrete orbital angular momentum (OAM) degree of freedom \cite{mair01,vaziri02,vaziri03,walborn04a,peeters07,pires10}, and a number of measurement devices have been developed to perform OAM measurements of single photons \cite{leach02,wei03,berkhout10}, as well as fractional-valued OAM projections \cite{oemrawsingh05,pors08,pors08b}. Given the vast possibilities of producing entangled photons and manipulating their spatial properties, it is interesting to determine the strengths and weaknesses of this system and identify what optical devices are necessary to perform a universal set of logic operations for investigations of quantum computation.   

\par
As advantages of the spatial variables of single photons for experimental investigations of CV algorithms we can mention:  (i) high quality entanglement is readily available with spontaneous parametric down-conversion sources and coincidence counting; (ii) diverse and robust quantum state engineering is possible using masks, gratings, spatial light modulators, and other diffractive devices, which allow for the production of non-Gaussian states and the implementation of non-Gaussian operations; (iii) photon losses in this case do not affect the fidelity of the quantum state or operation, but rather are discarded as null results;  (iv) contrary to quadrature detection, no local oscillator is necessary in the detection of transverse position and momentum of single photons. The transverse position and momentum are defined with respect to some transverse plane and the phase at the detection, which is relative to the phase at the source, can be varied through simple free propagation or using lenses.  
\par
In comparison to the usual quadrature variables of single modes, disadvantages of transverse spatial variables of single photons are: (a) the production of single or multiple photon states is probabilistic and post-selected; (b) the photon-photon coupling required for controlled operations is very weak;  (c) the single photon detection process is based on avalanche photodiodes, which have quantum efficiencies up to approximately 80\%, depending on the wavelength. At present, item (a) is due to the use of down-conversion sources for the production of single or multi-photon states.  Current efforts in the production of heralded single and multi-photon states may improve the situation.  In many cases the probabilistic production is not problematic, since typically only the coincidence detection events are considered.  Item (b) might be overcome using linear optical devices and post-selection, as has been done in the case of discrete quantum information processing with linear optics \cite{klm01,gkp01}.      
The above considerations suggest that the spatial degrees of freedom of multi-photon states may lend themselves well to the cluster state model of quantum computing, in which an initial entangled state is prepared, and sequential single-site measurements are performed.

\par
In this paper we show how to implement CV quantum logic gates using the spatial degrees of freedom of single photons, with the goal of encouraging future experimental investigations of CV quantum information processing with this system. 
We also discuss how in principle these gates can be used to create CV cluster states
in the transverse spatial degree of freedom of single photons, {\it i.e.}  the principal resource in the one-way
quantum computation  \cite{raussendorf01}. It is important to emphasize that our formalism
follows the traditional approach in quantum computation with any CV system where  the computational basis is given by the eigenstates associated with one of the canonical quadratures (transverse position or momentum of single photons in our case).
The fact that these are unphysical quantum states (not normalizable) forces one to use appropriate regularized 
Gaussian squeezed states as the computational basis in quantum protocols \cite{adcock09} or in the construction of cluster states for measurement-based quantum computation \cite{vanloock07, gu09}. 

Many properties of Gaussian cluster states have been investigated,  though its use as a scalable resource for universal for CV measurement-based quantum computation has not been demonstrated  for Þnite squeezing \cite{adcock09,eisert10,cable10}. In Ref. \cite{eisert10}, the authors analyze the effect of finite squeezing on the localizable entanglement (LE) \cite{verstraete05} between two general sites in a Gaussian graph state. They show that the LE decays exponentially with the distance between the sites in the graph, even if non-Gaussian (local) projective measurements are allowed. Although this result does not imply in the non-universality of Gaussian cluster states  with Þnite squeezing, it gives strong evidences of it and points toward the necessity of using non-Gaussian resources for Universal measurement-based quantum computation. The results of Ref. \cite{cable10}, though slightly more optimistic than those of Ref. \cite{eisert10}, reinforce this view. In this regard, the ability to easily manipulate the spatial variables of single photons could be of great interest for the study of spatially non-Gaussian cluster states.

The paper is outlined as follows.  In section \ref{sec:Spatial DOF} we introduce an operator formalism which can be used to describe the spatial degrees of freedom of single photons. In section \ref{sec:single} we show how the usual single-mode CV quantum gates can be performed with linear optics elements alone.  These gates, along with the two-photon gate proposed in section \ref{sec:two}, provide a universal set of gates for quantum information processing. Section \ref{sec:cluster} is devoted to the discussion of CV cluster computation with spatial degrees of freedom. Conclusions are presented in section \ref{sec:conc}. 

\section{Operator formalism for spatial degrees of freedom of single photons}
\label{sec:Spatial DOF}

We consider here the spatial degrees of freedom (DOF) of single photon states in the paraxial approximation, as studied previously by several authors \cite{caetano03,nogueira04a,walborn03b,walborn03a,howell04,dangelo04,almeida05,walborn06,abouraddy07,yarnall07,tasca08,walborn08,tasca09,tasca09b,gomes09b,gomes09a,pires09c,lvovsky09}.  It has been shown that  it is possible to establish a complete isomorphism between the Hilbert space describing the transverse spatial DOF of single photons and the Hilbert space associated with the non-relativistic quantum states of single point particles in a two dimensional space \cite{tasca09,lvovsky09}.
In this respect, one may think of the transverse field distribution of a single photon as a wave function in the formalism of first quantization. Thus, this wave function represents the probability amplitude for the detection of the single photon at a certain position in the transverse plane. 
\par
As the total  Hilbert space describing the spatial DOF of single photons is the tensor product of the Hilbert spaces associated with the two orthogonal transverse spatial directions, we will only describe the formalism in one spatial dimension.  Let us define the canonical dimensionless variables representing transverse position and transverse wave-vector of the single photon as $x$ and $p$, respectively.  Hilbert space $\mathcal{H}$ is spanned by the bases  
$\{\ket{j}\}$, where $j=x$ or $p$.  An arbitrary pure state of $\mathcal{H}$ is written as $\ket{\psi}= \int \psi(j)\ket{j} dj$, where $\langle j\ket{\psi}=\psi(j)$ is the transverse wave-function in the $j$ representation. For $j=x$ we have $\psi(j)=\mathcal{W}(x)$, which is the wave function in position representation or the transverse spatial distribution of the single photon field. The wave-function in wave-vector representation is the angular spectrum of the photon field and is obtained setting $j=p$ such that $\psi(j)=\mathcal{V}(p)$. 
The free evolution of the wave function associated with the transverse spatial DOF of a paraxial single photon is described, in the position representation, 
by the paraxial wave equation \cite{marcuse}:

\begin{equation}
\left(\frac{\partial^2}{\partial x_d^2} + 2ik \frac{\partial}{\partial z}\right)\mathcal{W}(x_d)=0,
\label{eq:wave}
\end{equation}
which is a Schr\"odinger-type equation analogous to that of a free massive particle with one DOF. Note that the coordinate of the paraxial direction of propagation $z$ plays the role of time and $\lambda$ the role of Planck's constant. Here $k =2 \pi/\lambda $ is the wave number and $x_d = d \cdot x$ is a dimensional position variable, and $d$ has dimension of length.  Similarly, we will use a dimensional momentum variable $p_d = p/d$, the cannonical conjugate of $x_d$.  Eq. \eqref{eq:wave} has the general solution

\begin{equation}
\mathcal{W}(x_d,z)=\exp\left(\frac{i z}{2k}\frac{\partial^2}{\partial x_d^2} \right)\mathcal{W}(x_d,0).
\label{eq:wavesol}
\end{equation}
\par
For convenience, we will adopt an operator formalism for the transverse spatial variables \cite{stoler81,marcuse}.  In terms of operators and ket vectors, Eq. \eqref{eq:wavesol} can be written as
\begin{equation}
\ket{\mathcal{W}(x_d,z)}=\exp\left(\frac{-i z}{2k}\oper{p}_d^2 \right)\ket{\mathcal{W}(x_d,0)},
\label{eq:wavesolket}
\end{equation}
where we have identified the dimensional momentum operator as $\oper{p}_d \rightarrow -i \partial/\partial x_d$.  The operator 

\begin{equation} \label{eq:P}
\oper{P}_z=\exp({-i z}\oper{p}_d^2/2k),
\end{equation}
describes free-space propagation from the origin to position $z$.  The operator formalism and ket vectors introduced by Stoler \cite{stoler81} and Marcuse \cite{marcuse} describe the transverse spatial properties of an electromagnetic field.  Here, we employ the same formalism to describe single-photon fields. Thus, we emphasize that $\ket{\mathcal{W}(x_d, z)}$  in Eq. \eqref{eq:wavesolket} represents the quantum state of a quasi-monochromatic single photon 
in the position representation of the transverse spatial DOF and $\ket{x}$ and $\ket{p}$ are ket vectors describing single photons in position or momentum eigenstates. 

\par
The wave-function evolution described by Eq. \eqref{eq:wavesolket}, associated with the free paraxial propagation of the single-photon field, is analogous to the evolution of a free particle. Other Hamiltonians can be implemented with the help of optical components such as lenses. In the operator formalism the action of a lens is given by \cite{stoler81}
\begin{equation}
\oper{L}_f = \exp \left(\frac{-ik}{2f}\oper{x}_d^2 \right),
\label{eq:L}
\end{equation}
where $f$ is the focal length.  In the next section, we will use the operators $\oper{P}_z$ and $\oper{L}_f$ to build a set of logic gates.  Here we assume that the typical dimensions of the
optical system used is sufficiently large so that
we can safely neglect diffraction effects, such as with the edges of lenses, for example.  Thus, the size of the optical system provides a constraint which limits the size of the single photon field.  This limits the amount that one can focus the field in the conjugate variable, and also the amount of information that can be encoded in the transverse profile.  This is equivalent to an energy constraint for intense fields which limits the amount of squeezing that can be performed. We note that, even with this constraint, it was possible to encode around 5 bits of information per photon in Ref. \cite{walborn06}.

At times it will be convenient to work with the dimensionless variables $x$ and $p$, which can always be obtained by introducing an appropriate scaling factor $d$.   
The operators $\oper{x}$ and $\oper{p}$ act on the elements of the position and wave-vector basis in the usual way: $\oper{x} \ket{x}=x\ket{x}$ and $\oper{p} \ket{p}=p\ket{p}$. The bases $\{\ket{x}\}$ and $\{\ket{p}\}$ are related via Fourier transform

\begin{equation}
 \ket{ x}  =  \frac{1}{\sqrt{2 \pi}}\int d p \, e^{-i  x \cdot p} \ket{p}, 
  \label{eq:ketx}
\end{equation}
\begin{equation}
\ket{p}  =  \frac{1}{\sqrt{2 \pi}}\int d x \, e^{i x \cdot p} \ket{x},
 \label{eq:ketp}
\end{equation}
where $\langle x | p \rangle = \exp(ixp)/\sqrt{2 \pi}$. Note that Eqs. \eqref{eq:ketx} and \eqref{eq:ketp} imply that the operators $\oper{x}$ and $\oper{p}$ satisfy the canonical comutation relation $[\oper{x},\oper{p}]=i$.

\section{Single-photon gates}
\label{sec:single}

It has been shown for quadrature variables (see for example \cite{lloyd99}) that for the subclass of unitary transformations that correspond to Hamiltonians that are polynomial functions of the canonical operators of a continuous variable system, it is possible to define a universal set comprising a finite number of quantum logical gates of qu-modes.  Any computation within this subclass of transformations can be decomposed into a finite number of applications of the gates in this universal set, which consists of a two-mode interaction, such as a beam splitter, and a set of single-mode quantum gates. The single-mode gates in the universal set are phase-space displacements and rotations, squeezing, and any operation that is at least of third-order in the canonical variables.  We will now show how to implement these CV ``qu-mode" operations in terms of the spatial DOF of photons, and discuss possible experimental implementations.  The analog to the single-mode gates will be single-photon gates, and a two-photon gate is the analog to the two-mode gate.  The two-photon gate will be described in section \ref{sec:two}.  As is customary, we will consider the position basis $\{\ket{x}\}$ as the computational basis. 
As mentioned above, these eigenstates are not normalizable, and thus unphysical, in the same sense as quadrature states with infinite squeezing.  
However, they can be approximated by physical states with very small variance \cite{lloyd99}. 

\subsection{Single-lens system}  

\begin{figure}
\begin{center}
\includegraphics[width=5cm]{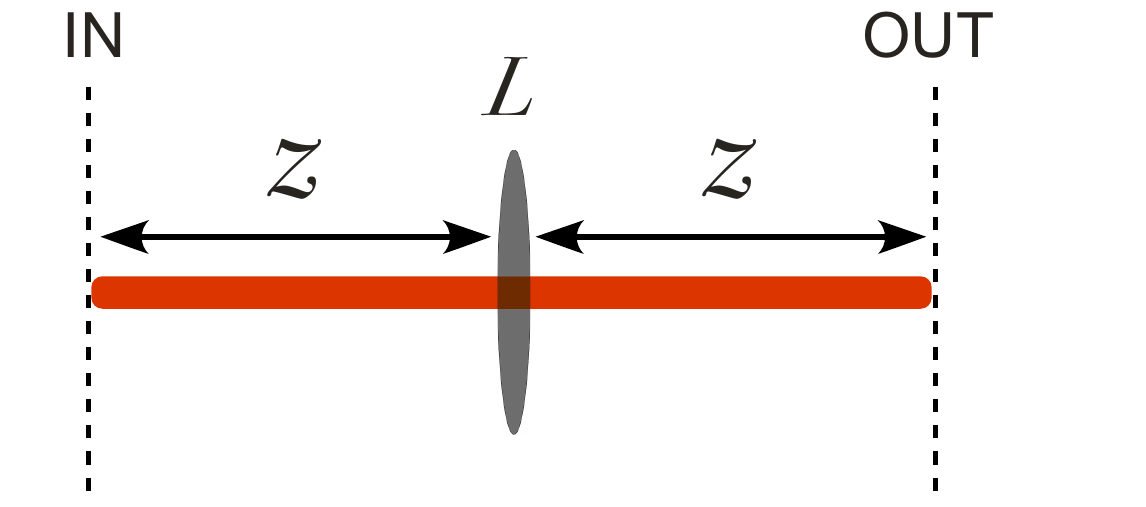}
\caption{ (color online) Optical setup for the implementation of single photon gates.  The single-photon field propagates from the input plane (IN) on the left to the output plane (OUT) on the right.  $L$ is a lens with focal length $f$ and $z$ is the distance of propagation before and after the lens.}
\label{fig:lens}
\end{center}
\end{figure}
The majority of the necessary single-photon gates can be implemented with the single-lens system illustrated in Fig. \ref{fig:lens}.  The operator describing this system is given by $\oper{U}(z, f)= \oper{P}_z\oper{L}_f\oper{P}_z$.  Explicitely, 
 \begin{equation}\label{eq:U}
\oper{U}(z, f) = \exp\left(-\frac{i z}{2k}\oper{p}_d^2 \right) 
 \times \exp \left(-\frac{ik}{2f}\oper{x}_d^2 \right) \exp\left(-\frac{i z}{2k}\oper{p}_d^2 \right). 
\end{equation} 
The operators $\oper{x}_d$ and $\oper{p}_d$ under the action of $\oper{U}(z, f)$ evolve as 
\begin{equation}
\oper{U}^{\dagger}(z, f) \, \oper{x}_d  \, \oper{U}(z, f) = \left(1-\frac{z}{f}\right) \oper{x}_d + \frac{z}{k}\left(2-\frac{z}{f} \right) \oper{p}_d;
\label{eq:xevol}
\end{equation}  
\begin{equation}
\oper{U}^{\dagger}(z, f) \, \oper{p}_d \, \oper{U}(z, f) = \left(1-\frac{z}{f}\right) \oper{p}_d - \frac{k}{f} \oper{x}_d,
\label{eq:pevol}
\end{equation}  
which exactly match the classical  evolution of transverse position, $x_d$, and wave vector, $p_d$. The phase-space operators ($\oper{x}_d, \oper{p}_d$) of a CV quantum system evolve, for any evolution operator associated with a quadratic hamiltonian (a metaplectic operator), as a multiplication by the symplectic matrix associated with the linear classical evolution of these operators \cite{guillemin}. In the paraxial propagation of photons, this classical evolution is described by the ray optics matrix applied to the optical system, i.e.,
\begin{equation}\label{eq:sympletictransform}
\oper{U}^{\dagger}(z,f) 
\left(
\begin{array}{cc}
 \oper{x}_d \\
 \oper{p}_d   
\end{array}
\right)
\oper{U}(z,f) =\mbox{M}(z,f)
\left(
\begin{array}{cc}
 \oper{x}_d \\
 \oper{p}_d   
\end{array}
\right),
\end{equation}
where

\begin{equation}\label{eq:Matrix}
\mbox{M}(z,f)=
\left(
\begin{array}{cc}
 1- \frac{z}{f} & \frac{z}{k}\left(2-\frac{z}{f}\right) \\
 -\frac{k}{f} & 1- \frac{z}{f} 
\end{array}
\right).
\end{equation}

\subsection{Fourier transform}
One of the building blocks of the universal set of single-mode gates is the Fourier transform operation, which is a special case of phase-space rotation. The Fourier transform operator $\oper{F}$ is the CV analog to the Hadamard gate for qubits \cite{chuang00}. The Fourier transform gate can be performed on the transverse spatial DOF of photons by using the lens system shown in Fig. \ref{fig:lens} with  $z=f$. Input and output planes which are related by a Fourier transform sit in the focal planes of the lens with focal length $f$. In this case, Eqs. \eqref{eq:xevol} and \eqref{eq:pevol} become

 \begin{equation}
\oper{F}^\dagger \oper{x}_d \oper{F} = \frac{f}{k} \oper{p}_d;
\label{eq:xdevolF}
\end{equation}  

\begin{equation}
\oper{F}^\dagger \oper{p}_d \oper{F} = -\frac{k}{f} \oper{x}_d.
\label{eq:pdevolF}
\end{equation}  
To change to dimensionless variables, we choose $d=\sqrt{f/k}$, which leads to 
\begin{equation}
\oper{F}^\dagger \oper{x} \oper{F} = \oper{p};
\label{eq:xevolF}
\end{equation}  
\begin{equation}
\oper{F}^\dagger \oper{p} \oper{F} = - \oper{x}.
\label{eq:pevolF}
\end{equation} 

The application of $\oper{F}$ takes position eigenstates to wave-vector eigenstates (and vice versa). Using Eqs. \eqref{eq:xevolF} and  \eqref{eq:pevolF} we get

\begin{equation} \label{eq: F on x and q}
\oper{F}\ket{p}  = \ket{x=p}; 
\end{equation}
\begin{equation}
  \oper{F}\ket{x}  = \ket{p=- x},
\end{equation}
where, for example, the ket $\ket{x=p}$ is an eigenvector of the position operator with eigenvalue $p$. This reflects the fact that after the application of the Fourier operation the dimensionless transverse position in the output plane corresponds to the dimensionless momentum in the input plane and vice versa (up to a sign). Thus, after the Fourier transformation,  the angular spectrum of the photon field in the input plane is mapped onto the position wave function of the photon field in the output plane:

\begin{equation}
\mathcal{W}_{out}(x)=\bra{x}\oper{F} \ket{\psi_{in}}= \bra{p} \psi_{in} \rangle= \mathcal{V}_{in}(p) .
\end{equation}

\subsection{Phase space rotations}

The operator describing rotations in the phase space of a CV quantum system with one DOF can be written as

\begin{equation}
\label{EQ:FRFToperator}
\oper{F}_{\theta} \equiv e^{i \theta/2}
\exp{\left(-i\theta\frac{\oper{x}^2+\oper{p}^2}{2}\right)}.
\end{equation}
It's action on the pair of conjugated operators $\oper{x}$ and $\oper{p}$ is 
\begin{equation} \label{EQ:FRFT rotations}
 \oper{F}_{\theta}^{\dagger}\,\oper{x}\, \oper{F}_{\theta} = \cos \theta \oper{x} + \sin\theta\oper{p}, 
 \end{equation}
 \begin{equation}
   \oper{F}_{\theta}^{\dagger} \,\oper{p} \, \oper{F}_{\theta} = - \sin\theta\oper{x} + \cos\theta\oper{p}. 
\end{equation}

For the spatial DOF of a single-photon field, rotations in phase space can be implemented using the Fractional Fourier Transform (FRFT) \cite{lohmann93,pellat-finet94,ozaktas01}.  It is possible to implement the FRFT on spatial DOF with free space propagation alone \cite{pellat-finet94} or with linear optical systems composed of free propagation and lenses \cite{lohmann93}.  Using the lens system shown in Fig. \ref{fig:lens}, we set the distance of propagation to $z_{\theta}=2 f \sin^2 \theta/2$ and define $f^\prime = f \sin \theta$, which is the fractional focal length. Now, introducing the scaling parameter $d = \sqrt{f^{\prime}/k}$ on the phase space operators, it is possible to describe this optical system as a fractional Fourier transform operator: $\oper{F}_{\theta}=\oper{U}(z_{\theta}, f)$. It is worth emphasizing that the appropriate scaling factor $d = \sqrt{f^{\prime}/k}$ is crucial for the description of the field transformation through the optical system $\oper{U}(z_{\theta}, f)$ as a FRFT (i.e. a rotation in phase-space). Here $\theta$ is the parameter which defines the ``order" of the FRFT (or the rotation angle in phase space). 

Note that the Fourier transform operation defined previously is the particular case where $\theta = \pi/2$.  This optical setup is able to implement rotations between $0<\theta<\pi$.  Rotations of angles larger than or igual to $\pi$ can be implemented with two or more composite FRFT systems, as long as the $f^{\prime}$ parameter is the same in every consecutive FRFT system.   It is straightforward to show that the matrix element $F_{\theta}(x,x^\prime) = \langle x|\oper{F}_{\theta}|x^\prime \rangle$ is given by  
\begin{equation}
F_{\theta}(x, x^\prime) =  A_{\theta} \exp \left [ i \frac{\cot \theta}{2}(x^2+{x^{\prime}}^2) \right ] \exp 
\left({-i\frac{x x^\prime}{\sin \theta}}\right),
\label{eq:E2}
\end{equation}
where $A_{\theta}=\sqrt{i \exp(i \theta)/2 \pi |\sin \theta|}$.  $F_{\theta}(x, x^\prime)$ is the usual FRFT kernel in optics, and is also equivalent to the propagator of the simple quantum harmonic oscillator (of frequecy $\omega$) with the parameter $\theta = \omega(t-t_0)$ \cite{ozaktas01}.      

\subsection{Squeezing}
\begin{figure}
\begin{center}
\includegraphics[width=6cm]{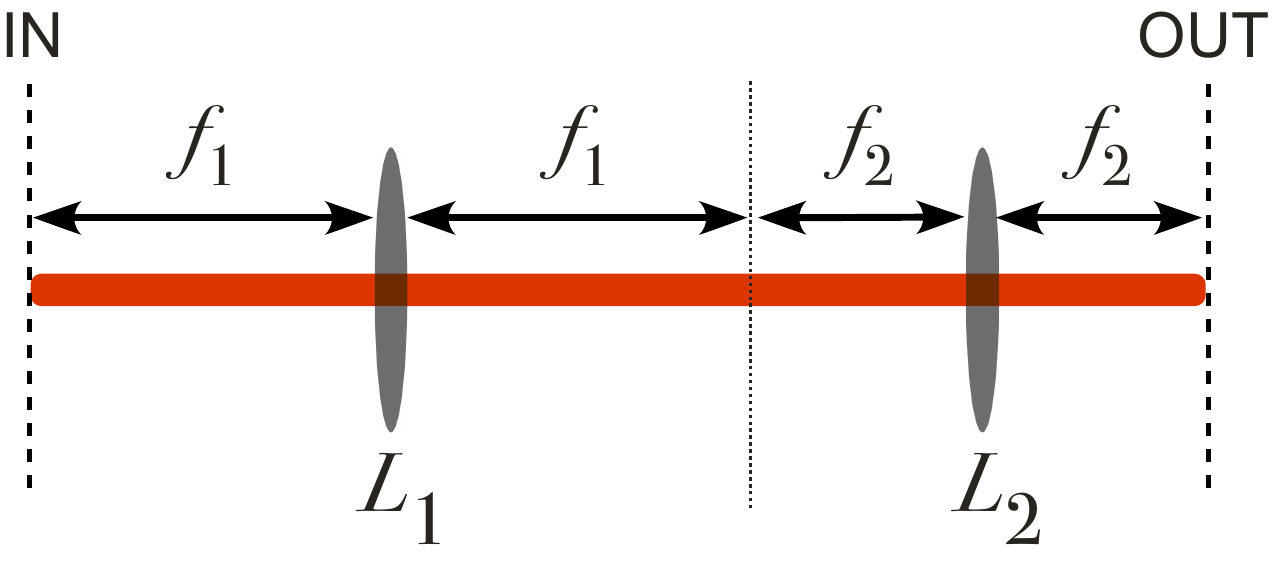}
\caption{ (color online) Optical setup for the squeezing gate. Two lenses $L_1$ and $L_2$ with focal lengths $f_1$ and $f_2$, respectively, are placed in a confocal setup. }
\label{FIG:Squeezing}
\end{center}
\end{figure}

In CV systems the operator
\begin{equation}
{\oper{S}} = \exp \left [-\frac{i}{2}\ln r  \left ( \oper{x}\oper{p} +\oper{p} \oper{x} \right ) \right ]
\end{equation}
corresponds to a squeezing operation with the squeezing parameter $r$. The action of this operator on the pair of conjugated variables is
\begin{subequations}
 \label{EQ:squeezing}
\begin{equation}
 {\oper{S}}^{\dagger}\,\oper{x}\, {\oper{S}} = r \oper{x}, 
\end{equation}
\begin{equation}
{\oper{S}}^{\dagger} \,\oper{p} \,{\oper{S}} = \frac{1}{r} \oper{p}. 
\end{equation}
\end{subequations}
In principle, the squeezing gate for the transverse spatial DOF is simply the focusing by a lens. A collimated beam, focused by a convergent lens, assumes its waist at the focal plane. The width of the beam at the waist will depend on the focal length and on parameters such as the position and waist before the lens. Nevertheless, this beam may diverge drastically after passing through the focal plane, especially in the case of large squeezing. To construct an  optical squeezing system which preserves the collimation of the beam we make use of the system shown in fig. \ref{FIG:Squeezing}. Two lenses of focal lengths $f_1$ and $f_2$ are arranged in a confocal configuration, resulting in an optical system described by the operator $\oper{U}_{f_1}\oper{U}_{f_2}\equiv\oper{U}(z=f_1, f_1)\oper{U}(z=f_2, f_2)$. This system implements two consecutive Fourier transform optical gates $\oper{F}_{\frac{\pi}{2}}$ for the dimensional phase-space operators $(\oper{x}_d, \oper{p}_d)$, each one with a different focal length parameter. Direct application of Eqs. \eqref{eq:xevol} and \eqref{eq:pevol} shows that the complete transformation performed by this system is
\begin{equation}
\oper{U}^\dagger_{f_2} \oper{U}^\dagger_{f_1} \, \oper{x}_d \, \oper{U}_{f_1}\oper{U}_{f_2} =- \frac{f_2}{f_1} \, \oper{x}_d; \end{equation}  
\begin{equation}
\oper{U}^\dagger_{f_2} \oper{U}^\dagger_{f_1} \, \oper{p}_d \, \oper{U}_{f_1}\oper{U}_{f_2}=- \frac{f_1}{f_2} \oper{p}_d,
\end{equation}  
which is the squeezing gate with squeezing parameter $r=f_2/f_1$.  Thus, we have $\oper{S} = \oper{U}(f_1, f_1)\oper{U}(f_2, f_2)$.  We note that this transformation is a squeezing operation regardless of the scaling parameter $d$. We can still keep $d=\sqrt{f^{\prime}/k}$, which is necessary for the application of the FRFT gate, as our scaling parameter. It is essential to keep the same scaling parameter along the entire computation in order to be able to describe all the operations in the same phase space.
 Note that this gate is equivalent to an imaging system with magnification, and the minus sign accounts for the inversion of the image, which could be undone with a phase space rotation. For $f_1=f_2$ we have as output the inverted image of the input field with unit magnification. The application of the squeezing optical gate (fig. \ref{FIG:Squeezing}) with the input plane corresponding to position space, squeezes the transverse operators according to Eqs. (\ref{EQ:squeezing}). Its application when the input plane corresponds to momentum space (i.e. after the application of a Fourier transform
on the input field), will squeeze the operators in the opposite way, dividing $\oper{x}$ by $r$ and multiplying $\oper{p}$ by $r$. 

\subsection{Pauli gates}
The CV analog to the usual Pauli gates $\sigma_z$ and $\sigma_x$ are defined as 
\begin{equation} \label{eq:pauliX}
{\oper{X}}(t) = \exp (-i t  \oper{p}),
\end{equation}
and
\begin{equation}\label{eq:pauliZ}
{\oper{Z}}(s) = \exp (i s  \oper{x}) ,
\end{equation}
where $s$ and $t$ are displacement parameters. The action of these gates is to displace the position and momentum eigenstates as ${\oper{X}}(t)\ket{x} \longrightarrow \ket{x+t}$ and ${\oper{Z}}(s)\ket{p} \longrightarrow \ket{p+s}$.  The elements of the computational basis $\{\ket{x}\}$ are the eigenstates of ${\oper{Z}}(s)$ with eigenvalues $e^{i s \cdot x}$ while the elements of the conjugate basis $\{\ket{p}\}$ are  eigenstates of ${\oper{X}}(t)$ with eigenvalues $e^{-i t p}$.  The evolution of the phase-space operators under the action of the CV Pauli operators is
\begin{subequations}
\label{eq:EvolutionpauliX}
\begin{equation} 
\oper{X}^{\dagger}(t) \, \oper{x} \, {\oper{X}}(t) = \oper{x} + t \, ;
\end{equation}
\begin{equation}
\oper{X}^{\dagger}(t) \, \oper{p} \, {\oper{X}}(t) = \oper{p} \, ,
\end{equation}
\end{subequations}
and
\begin{subequations}
\label{eq:EvolutionpauliZ}
\begin{equation}  
\oper{Z}^{\dagger}(s) \, \oper{x} \, {\oper{Z}}(s) = \oper{x} \, ;
 \end{equation}
\begin{equation}
\oper{Z}^{\dagger}(s) \, \oper{p} \, {\oper{Z}}(s) = \oper{p} + s \, , 
\end{equation}
\end{subequations}
Eqs. \eqref{eq:EvolutionpauliX} and \eqref{eq:EvolutionpauliZ} imply that the CV Pauli gates will displace the wave function of the single photon state. For position displacement we have
\begin{equation}\label{eq:EvolutionpauliW}
\mathcal{W}_{out}(x)  = \bra{x} \oper{X}(t) \ket{\psi_{in}} = \bra{x-t} \psi_{in}\rangle = \mathcal{W}_{in} (x-t).
\end{equation}
For wave vector displacement, we have
\begin{equation}\label{eq:EvolutionpauliV}
\mathcal{V}_{out}(p)  = \bra{p} \oper{Z}(s) \ket{\psi_{in}}= \bra{p-s} \psi_{in}\rangle = \mathcal{V}_{in} (p-s).
\end{equation}

Position displacements of the transverse wave function of paraxial photons could be implemented with the help of a pair of reflecting or refracting optical elements such as mirrors or prisms. The optical elements should be positioned in the path of the optical field in order to displace it with respect to the axis of propagation. For momentum displacements, one would have first to map the momentum distribution onto position space, and then apply the displacement.  
This scheme, nevertheless, might be difficult to implement for small displacements (in comparison to the width of the transverse distribution).

A more realistic scheme makes use of a basic property of the Fourier transform, namely the shift theorem \cite{goodman96}:  translation of one variable corresponds to a linear phase shift in its conjugated variable. The angular spectrum of the position displaced wave function Eq. \eqref{eq:EvolutionpauliW} is
\begin{equation}\label{eq:shiftW}
\mathcal{V}_{out}(p)  = \bra{p} \oper{X}(t) \ket{\psi_{in}}=e^{-itp} \bra{p} \psi_{in}\rangle = e^{-itp} \, \mathcal{V}_{in} (p). 
\end{equation}
The wave function in position representation relative to the displaced angular spectrum Eq. \eqref{eq:EvolutionpauliV} is 
\begin{equation}\label{eq:shiftV}
\mathcal{W}_{out}(x)  = \bra{x} \oper{Z}(s) \ket{\psi_{in}}= e^{isx} \bra{x} \psi_{in}\rangle = e^{isx} \, \mathcal{W}_{in} (x).
\end{equation}

\begin{figure}
\begin{center}
\includegraphics[width=4cm]{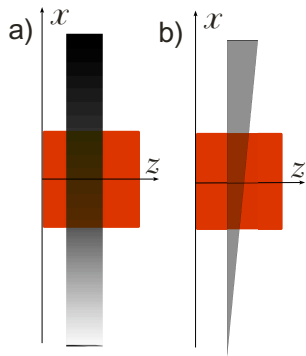}
\caption{ (color online) Linear phase shifter. a) Representation of the graded-refractive-index LPS. It is composed of a transparent optical plate with width $l$ which refractive index $n(x)$ is a linear function of the transverse position $x$. b) Wedge-shaped LPS. Transparent optical plate with constant refractive index and thickness that changes linearly with transverse position.}
\label{FIG:LPS1}
\end{center}
\end{figure}

Thus,  displacement in one variable can be achieved by inserting a linear phase shift on its conjugate variable. For example, a phase shift $s \cdot x$ in position domain corresponds to a displacement $s$ in wave-vector domain. In practice, this can be achieved with the spatial DOF of photons  by inserting an optical element which introduces a linear phase shift on the transverse field distribution.  We call this element a ``linear phase shifter" (LPS). It could be, for example, a transparent plate with length $l$ and transverse linear modulation of its refractive index: $n(x_d)= n_0 x_d$  (see Fig. \ref{FIG:LPS1}-a). In this case the optical path varies linearly with the transverse position, so that the displacement in the conjugated dimensionless variable will be $  s = n_0 l d$. Another possibility is a wedge-shaped plate with constant refractive index  (see fig. \ref{FIG:LPS1}-b).  In this case the linear modulation of the width of the phase-plate makes the optical path vary linearly with the transverse position.  Perhaps the most promising method with current technology is to produce the linear phase shift with a programmable spatial light modulator. 

\begin{figure}
\begin{center}
\includegraphics[width=6cm]{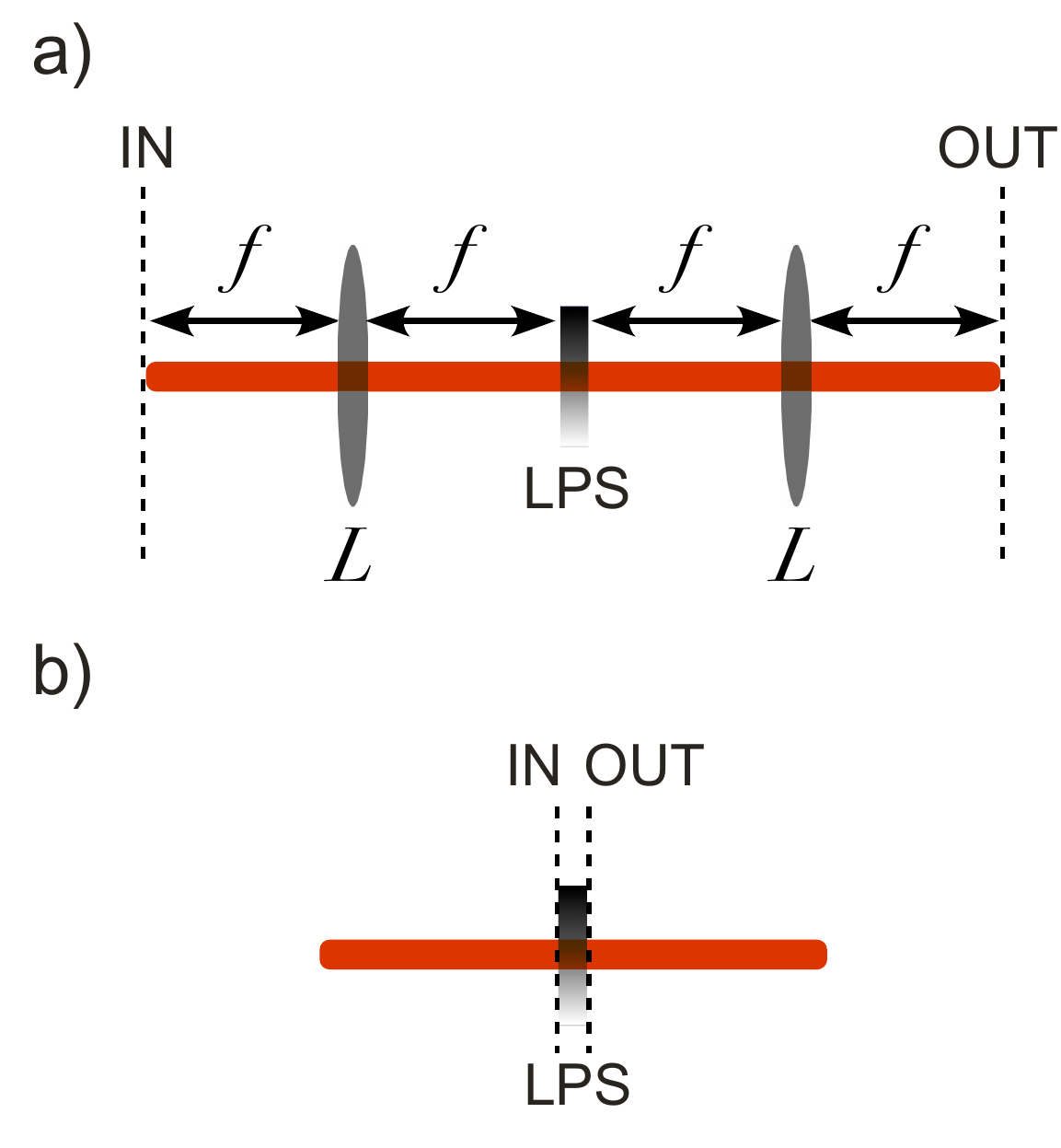}
\caption{ (color online) a) Optical setup for the implementation of the position displacement gate. The first Fourier transform system maps the angular spectrum of the input state to the transverse field distribution at the forward focal plane, where a linear phase shift is introduced. The second Fourier system maps the phase-shifted field distribution back to wave vector space, resulting in a displaced position distribution. b) Optical setup for the implementation of wave-vector displacement gate. A LPS is applied on the initial field distribution and thus displaces its angular spectrum.}
\label{FIG:LPS2}
\end{center}
\end{figure}

The application of the LPS corresponds to the aplication of the $\oper{Z}(s)$ CV Pauli gate. Displacements of the position or momentum coordinates can be achieved by inserting the LPS respectively in the Fourier or imaging plane of the target photon field. In this way, displacement gates will in general involve  LPS's and Fourier transforms. Fig. \ref{FIG:LPS2}(a) shows the implementation of a displacement in position space. A Fourier transform gate maps the angular spectrum of the photon field onto position space. The LPS is then applied to the field distribution at the focal plane of the first lens and another Fourier transform gate is performed to map the initial position distribution (now displaced) back to position space. In terms of the operator formalism, the sequence of operations necessary to implement a displacement of the position wave function is 
\begin{equation} \label{eq:X=FZF}
\oper{X}(t=s)=\oper{F}^{\dagger} \, \oper{Z}(-s) \, \oper{F},  
\end{equation}
and can be proven using Eqs. \eqref{eq:xevolF} and \eqref{eq:pauliZ}. 

The Hermitian conjugate of the Fourier transform gate (or the inverse Fourier transform gate) involved in the CV Pauli gate $\oper{X}(s)$ expressed in Eq. \eqref{eq:X=FZF} can be implemented with three consecutives Fourier transform gates: $\oper{F}^{\dagger}=\oper{F}^3$, or a single Fourier transform and coordinate inversion.  Thus, the optical setup illustrated in Fig. \ref{FIG:LPS2} a)  implements the sequence $\oper{F} \, \oper{Z}(-s) \, \oper{F}$, which displaces the wave function as 
\begin{equation}
\mathcal{W}_{in}(x) \rightarrow \mathcal{W}_{out}(x) = \mathcal{W}_{in}(-x-s),
\end{equation}   
where the minus sign multiplying the position coordinate is due to the inversion caused by the two Fourier transform gates.  If it is necessary to recover the initial orientation, another imaging system $\oper{F}^2$ can be applied at the end of the operation. Fig. \ref{FIG:LPS2} b) shows the optical setup required to perform a displacement in the momentum domain. It consists of an LPS, which is applied directly in the path of the photon field, and shifts its phase in position space, according to Eqs. \eqref{eq:EvolutionpauliV} and \eqref{eq:shiftV}. 

We finally note that the choice of the scaling parameter $d$ will affect the amount of displacement introduced by the optical systems described in FIGs. \ref{FIG:LPS2}(a) and \ref{FIG:LPS2}(b) but, once its value is fixed, it is always possible to describe the field transformation through these optical systems as a CV Pauli operation. The choice of the scaling paramenter as $d=\sqrt{f^{\prime}/k}$ is mandatory for the phase-space rotation gate, and thus might be adopted for the CV Pauli gates.

\subsection{Higher-order gates}
One advantage of the spatial DOF of photons is that higher order gates of the form 
\begin{equation}
{\oper{B}}(n,\alpha) = \exp (i \alpha \oper{x}^n),
\end{equation}
can be performed by simply modifying the spatial profile of the field.  Arbitrary modulation of the transverse profile of an optical beam can be made with a spatial light modulator (SLM). This device can work by transmission or reflection, and can be adjusted to perform an arbitrary phase modulation on the transverse profile of the transmitted or reflected beam.   The phase modulation can be applied in position space by directly placing the SLM on the path of the beam.  Phase modulations in momentum space can be performed with the help of an additional Fourier transform gate. Note that the SLM can also apply linear or quadratic phases as well, thus allowing for the implementation of all single-photon gates discussed above.  

  Finally, it is important to remark that errors in the $z$ positioning of the optical elements (SLM or lenses) translates into an error in the single photon gates. However, if the Rayleigh length associated with the single photon sources is large, error in the $z$ positioning of an optical element leads to a negligible error in the implementation of the single photon gates.  Also, there is a limit to the resolution of position detection that can be achieved using an array of photodiodes, such as in an ICCD camera.  The spatial resolution of the transverse position is limited to the size of the pixels in the ICCD camera, which have typical dimensions of a few micrometers. Nevertheless, one can always magnify the transverse distribution of a single photon with an imaging system. The magnification could be applied in order to achieve the desired degree of precision in the transverse position discrimination, and would be limited ultimately by the size of the array of photodiodes, which is approximately $1$cm in most commercial ICCD cameras.

\section{Two-photon gates}
\label{sec:two}
\par
A general experimental difficulty for quantum information processing with light is the need for non-linear optical processes.   
For example, universal quantum computation with CV's defined in the quadrature variables of single modes, a  Kerr-type non-linearity is required to realize higher-order quantum gates \cite{lloyd99}.  On the other hand, for universal quantum computation with photonic qubits, optical non-linearity is needed to perform controlled logic operations of two or more qubits encoded in different photons \cite{milburn89}.  In Ref. \cite{klm01}, it was shown that this non-linearity could be introduced using only linear optical elements and post-selection.  This has allowed for the proof-of-concept realization of a number of multi-photon gates \cite{obrien03,langford05,kiesel05,okamoto05,lanyon08}. 
\par
The one-way model of quantum computation \cite{raussendorf01} is an interesting paradigm from the point of view of photonic quantum computing, since it allows for the multi-photon gates to be moved offline.  In the one-way model applied to this context, multi-photon gates would be used to construct an initial entangled state.  Processing then occurs through single-photon measurements and classical feed forward.  In this way, one can replace deterministic multi-photon gates with probabilistic entangling gates.  Smaller scale investigations even allow for the use of probabilistic sources of entangled photons.             
\par
For CV spatial DOF of photons, single photon gates to arbitrary order are implemented only with linear optical elements.  However, a photon-photon interaction would be required for two-photon CV gates such as a controlled-phase gate \cite{braunstein05}. Two-photon (or multi-photon) gates are required, whether for direct quantum
computation or for the construction of cluster states in the one-way model.
 Even though the photon-photon interaction is weak, small scale implementation of this sort of operation should be possible with current technology, and efficiencies might be improved in the future with the further development of photonic technology.   We note that investigations using the one-way model might take advantage of the spatial entanglement produced from spontaneous parametric down-conversion, which will be discussed in Section \ref{sec:cluster}. 
        \par
To illustrate the type of two-photon interaction required, we will outline a proof of principle method for entangling
the spatial variables of single photons using a nonlinear
four-wave mixing (4WM) interaction.   The Hamiltonian describing the 4WM process reads
\begin{equation}
 \oper{B}_{\mathrm{4WM}} \propto \int d\vect{\rho}\int dz[\oper{E}_{1}^{(+)}\oper{E}_{2}^{(+)}\oper{E}_{s}^{(-)}\oper{E}_{i}^{(-)}] + h.c. ,
\label{E2}
\end{equation}
where $(\oper{E}_{1},\oper{E}_{2})$ and  $(\oper{E}_{s},\oper{E}_{i})$ are the field operators for the two input and two output fields, respectively. The electric field operator in the paraxial approximation is given by \cite{mandel95}
\begin{equation}
{{\oper{E}}}(\vect{\rho},t) \propto e^{i(kz-\omega t)}\int d\vect{q} \oper{a}(\vect{q})e^{i\vect{q}\cdot \vect{\rho}} e^{-iq^{2}z/2k},
\label{E1}
\end{equation}
where $\vect{\rho}=(x, y)$ and $\vect{q}=(q_x, q_y)=(p_x, p_y)/\hbar$ are the two-dimensional transverse components of the position vector $\vect{r}$ and wave vector $\vect{k}$. The operator $\oper{a}(\vect{q})$ annihilates a photon with transverse wave vector \vect{q}.

Substituting the field operator (\ref{E1}) into the Hamiltonian (\ref{E2})  and integrating over $z$ and $\vect{\rho}$ one obtains
\begin{align}
\oper{B}_{\mathrm{4WM}} \propto   & e^{-i\Delta \omega t}\int d\vect{q}_{1}d\vect{q}_{2}d\vect{q}_{s}\sinc[(\Delta k -\Delta q^{2})L]   \times \nonumber \\
& \oper{a}(\vect{q}_{1})\oper{a}(\vect{q}_{2})\oper{a}^{\dagger}(\vect{q}_{s})\oper{a}^{\dagger}(\vect{q}_{1}+\vect{q}_{2}-\vect{q}_{s}) + h.c.
\end{align}
where $\Delta k = k_{1}+k_{2}-k_{s}-k_{i}$, $\Delta \omega = \omega_{1}+\omega_{2}-\omega_{s}-\omega_{i}$, and $\Delta q^{2}=q_{1}^{2}/2k_{1}+q_{2}^{2}/2k_{2}-q_{s}^{2}/2k_{s}-q_{i}^{2}/2k_{i}$, and  it has been assumed that the 4WM medium is large in the transverse directions and of length $L$ in the longitudinal direction.  Let us consider that frequency filters are used so that $\omega_1=\omega_i$ and $\omega_2=\omega_s$, which gives $\Delta \omega =0$ and $\Delta k=0$.    
If the length of the non-linear medium $L$ is small, the sinc function is practically constant and the Hamiltonian can be written as
\begin{align}
\oper{B}_{\mathrm{4WM}} \propto  & \int  d\vect{q}_{1}d\vect{q}_{2}d\vect{q}_{s}\oper{a}(\vect{q}_{1})\oper{a}(\vect{q}_{2}) 
 \oper{a}^{\dagger}(\vect{q}_{s})\oper{a}^{\dagger}(\vect{q}_{1}+\vect{q}_{2}-\vect{q}_{s}) \nonumber \\ 
 & + h.c.  
\label{E6}
\end{align}

Let us now return to the one-dimensional case.  
Application of the Hamiltonian $\oper{B}_{\mathrm{4wm}}$ to the initial momentum state $\ket{{q}_1}_1\ket{{q}_2}_2\ket{\mathrm{vac}}_s\ket{\mathrm{vac}}_i$ gives the output state of signal and idler photons
\begin{equation}\label{eq:EPR_4WM}
\ket{\psi} = N \int d{q} \ket{Q-{q}}_s\ket{{q}}_i,   
\end{equation}
 where $Q={q}_1+{q}_2$.  This state Eq. \eqref{eq:EPR_4WM} is a maximally entangled state or an EPR state.  We note that the 4WM process is not an implementation of a controlled-phase gate, since application of $\oper{B}_{\mathrm{4WM}}$ to the entangled state \eqref{eq:EPR_4WM} does not result in a product state.  Nevertheless, the 4WM process can be used to spatially entangle photons that are initially in a separable state.  In this fashion, we will show in the next section that it is possible to construct arbitrary entangled cluster states which can be used for CV quantum computation.          

\section{One-way computation with spatial DOF of photons}
\label{sec:cluster}
\begin{figure}
\begin{center}
\includegraphics*[width=6cm]{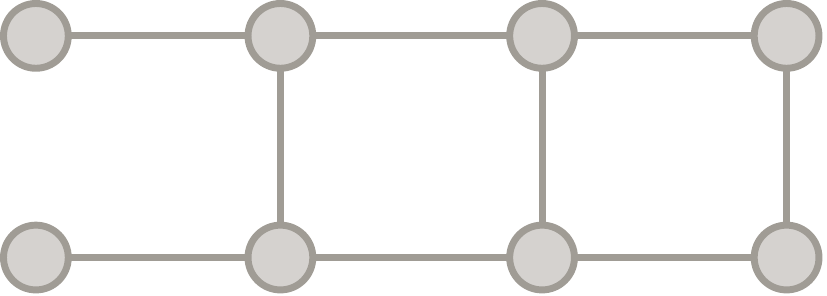}
\caption{\label{fig:clusterex} (color online)  Example of a cluster state.  The circles represent individual qu-modes, and the lines represent entanglement, created by a controlled-momentum displacement operation.}
\end{center}
\end{figure}

The one-way quantum computation model \cite{raussendorf01} extended to continuous variable sytems \cite{menicucci06,vanloock07,gu09} requires the production of an appropriate entangled cluster state, which is determined by the problem at hand. The basic recipe for the construction of CV cluster states consists of the preparation of momentum eigenstates with $p=0$ and implementation of controlled momentum displacement ($C\oper{Z}$) gates to entangle different nodes in the cluster.  An example of a cluster state is shown in Fig. \ref{fig:clusterex}.  Each node represents one DOF of a photon or mode, and the lines connecting each node represent entanglement, which is produced via the $\oper{CZ}$ gate.   The one-way computation is then carried out by performing sequential projective measurements on the individual nodes and classical feed forward of the measurement results.  For the power of universal computation, it is necessary to perform both gaussian and non-gaussian operations \cite{menicucci06}, just as in the quantum circuit model \cite{lloyd99}.  This can be achieved using a cubic-order (or higher) gate or by performing non-gaussian measurements.  We have shown above that the complete set of single-photon gates can be implemented on the spatial variables of single photons.  We will now show that it is possible in principle to build CV cluster states of photons.  To build the entangled cluster state in the spatial DOF of photons, one first prepares each photon in a zero momentum state $\ket{p=0}$, and then applies the gate sequence $\oper{F}_{j}\oper{B}_{\mathrm{4WM}ij}$ to each pair of photons connected by a line:
\begin{align}
\oper{F}_{j}\oper{B}_{\mathrm{4WM}ij}\ket{p=0}_i\ket{p=0}_j = & \oper{F}_j\int dp \ket{p}_i\ket{-p}_j \nonumber \\
= & \frac{1}{\sqrt{2\pi}}\iint dx dx^\prime e^{ixx^\prime}\ket{x^\prime}_i\ket{x}_j.   
\end{align}   
Applying the above operations to every link between nodes creates the CV cluster state.  

\subsection{Cluster state computing with two degrees of freedom per photon}
\begin{figure}
\begin{center}
\includegraphics*[width=8cm]{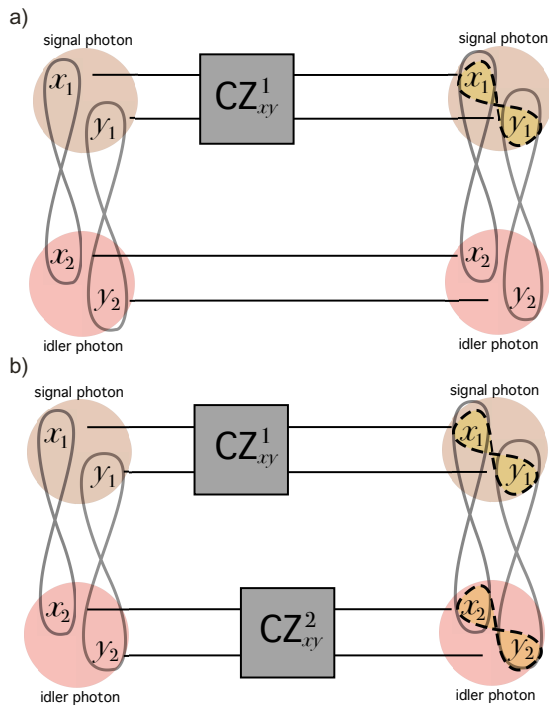}
\caption{\label{fig:cluster} (color online)  Circuit diagram illustrating the generation of CV's cluster states using two spatial degrees of freedom per photon of the two-photon state from SPDC. Evolution is from left to right.  The closed curves represent entanglement.  The CZ gates are performed on different spatial dimensions of the same photon.  Generation of a) linear cluster state  and b) ring cluster state.}
\end{center}
\end{figure}

One problem in quantum computing with single photons is the weak non-linearity, which is required for the two-photon gate in section \ref{sec:two}.  At present, the most immediate path for experimental investigation of CV quantum computing with spatial variables of photons is to produce entangled photon pairs directly using spontaneous parametric down-conversion (SPDC) \cite{walborn10a}, and exploit the bi-dimensionality of the transverse space.  This type of approach has been  employed in experimental investigations of qubit cluster states \cite{ceccarelli09}.  The spatial properties of the two-photon state from SPDC can be engineered in many ways \cite{torres03b,walborn03a,monken98a,gomes09b}. Under appropriate conditions, SPDC produces states in Einstein-Podolsky-Rosen-like states, which are equivalent to cluster states up to local transformations.   These cluster states can be produced by exploiting both transverse spatial DOF, that is in the $x$ and $y$ directions.  The SPDC state is \cite{monken98a,walborn03a}
\begin{equation}
\ket{\psi}_{\mathrm{SPDC}} = \int v(\vect{q}_1+\vect{q}_2)\gamma(\vect{q}_1-\vect{q}_2) \ket{\vect{q}_1}  \ket{\vect{q}_2} d\vect{q}_1d\vect{q}_2.
\label{eq:spdcstate}
\end{equation}
Typically, the function $\gamma$ is very large comparted to $v$, and can be approximated by unity.  The function $v$ is the angular spectrum of the pump laser beam \cite{monken98a}.  If the pump beam is well approximated by a plane wave, so that $v(\vect{q})\approx \delta(q_x)\delta(q_y)$, we have the product of two entangled states:
\begin{equation}
\ket{\psi}_{\mathrm{SPDC}} =  \int \ket{{q}_{x}}  \ket{-{q}_{x}} d{q}_{x} \int  \ket{{q}_{y}}  \ket{-{q}_{y}} d{q}_{y}.
\end{equation}
Applying a two-dimension Fourier transform  $\oper{F}_{2y}\oper{F}_{2x}$ to one of the photons, this state can be transformed to
 \begin{align}
\ket{\psi}_{\mathrm{SPDC}} \propto   & \iint e^{-i x_1 x_2}\ket{{x}_{1}}  \ket{{x}_{2}} d{x}_{1}d{x}_{2} \times \nonumber \\
&  \iint e^{-i y_1 y_2} \ket{{y}_{1}}  \ket{{y}_{2}} d{y}_{1}d{y}_{2},    
\end{align}
which is the product of two two-mode clusters states.  
The $x$ and $y$ DOF of a single photon $j$ can be entangled by applying a single-photon controlled $\oper{CZ}_{xy}^j$ gate:
\begin{equation}
\oper{CZ}^j_{xy} = \exp(i \oper{x}_j \oper{y}_j).  
\end{equation}
This operation can be easily implemented using a spatial light modulator, for example.  Applying the gate $\oper{CZ}_{xy}^1$  or  $\oper{CZ}_{xy}^2$ results in a linear cluster state, as in Fig. \ref{fig:cluster} a).  Applying $\oper{CZ}_{xy}^1\oper{CZ}_{xy}^2$ results in a ring cluster, as in Fig. \ref{fig:cluster} b).  
\par
Interesting investigations might exploit the angular spectrum transfer \cite{monken98a} of the pump beam to the two-photon state \eqref{eq:spdcstate}.  In this way, it is possible to produce non-Gaussian entangled states as in Ref. \cite{gomes09b}, which could serve as the CV cluster state.  We leave further investigation of generation of useful non-Gaussian states for future work.        

\section{Discussion and Conclusions}
\label{sec:conc}

The quantum correlations between the spatial degrees of freedom of photons produced in parametric down-conversion have been studied for more than twenty years. Even though the fundamental aspects of these correlations have been studied since the early nineties, their implications and possible applications in quantum computation and information are still subject of intense research.  On the one hand, the scalability of this system towards the implementation of reliable quantum computers is quite difficult with current technological resources. On the other hand, there have been a number of proof-of-concept experiments, particularly with polarization-entangled photons, which have strongly contributed to advance the understanding of quantum entanglement and its application in quantum information.  In this regard, we expect spatial DOF to allow for important investigations of CV quantum information processing.   
\par
Spatial degrees of freedom of photons are continuous variables, but the experiments are performed in the photon counting regime. In traditional quantum optics, it is very often the opposite. Even though the photon number is a discrete variable, the intensities of optical fields are converted into
electric currents, and their fluctuations are analyzed in terms of the continuous variable field quadratures. In this paper, we discussed the use of the spatial degrees of freedom of photons as a continuous variable associated to a single photon in a multimode field. The detection of a single photon, or two ``single" photons in coincidence post-selects the state of the field. This allows one to describe the single-photon multimode field with the same quantum formalism as a non-relativistic point particle.

Using this description of the single photon multimode field, we presented methods for the implementation of the basic logical gates and demonstrated that it is possible to produce entangled states of several photons, at least in principle. This allows for the construction of CV cluster states, which can be used for the one-way quantum computation architecture. We would like to stress the experimental simplicity for the implementation of many important operations. Rotations in phase space, Pauli gates and squeezing can be implemented with free propagation and lenses, and higher-order gates can be performed with a spatial light modulator. However, conditional operations are more complicated. Our proposal using four-wave mixing shows that it is possible to entangle two initially disentangled photons, which is equivalent to a conditional operation. Both the implementation of conditional gates and the production of clusters with this method are strongly limited by the low efficiency of the process. However, even at the low efficiency regime, it could be quite useful in preliminary investigations. Moreover, a notable technical evolution in the efficiency of non-linear optical processes, such as the development of periodically-poled non-linear crystals, has occurred in the last few years. 
\par
In conclusion, we have presented the basic elements necessary to the use of the spatial degrees of freedom of photons in quantum information and quantum computation.  We hope that this will open the door for novel experimental investigations of CV quantum information processing.

\begin{acknowledgements}
Financial support was provided by Brazilian agencies CNPq,
CAPES, FAPERJ, and the Brazilian Instituto Nacional de Ci\^encia e 
Tecnologia - Informa\c{c}\~ao Qu\^antica (INCT-IQ).
\end{acknowledgements}


\begin{thebibliography}{63}
\expandafter\ifx\csname natexlab\endcsname\relax\def\natexlab#1{#1}\fi
\expandafter\ifx\csname bibnamefont\endcsname\relax
  \def\bibnamefont#1{#1}\fi
\expandafter\ifx\csname bibfnamefont\endcsname\relax
  \def\bibfnamefont#1{#1}\fi
\expandafter\ifx\csname citenamefont\endcsname\relax
  \def\citenamefont#1{#1}\fi
\expandafter\ifx\csname url\endcsname\relax
  \def\url#1{\texttt{#1}}\fi
\expandafter\ifx\csname urlprefix\endcsname\relax\def\urlprefix{URL }\fi
\providecommand{\bibinfo}[2]{#2}
\providecommand{\eprint}[2][]{\url{#2}}

\bibitem[{\citenamefont{Nielsen and Chuang}(2000)}]{chuang00}
\bibinfo{author}{\bibfnamefont{M.}~\bibnamefont{Nielsen}} \bibnamefont{and}
  \bibinfo{author}{\bibfnamefont{I.}~\bibnamefont{Chuang}},
  \emph{\bibinfo{title}{Quantum Computation and Quantum Information}}
  (\bibinfo{publisher}{Cambridge}, \bibinfo{address}{Cambridge},
  \bibinfo{year}{2000}).

\bibitem[{\citenamefont{Lloyd and Braunstein}(1999)}]{lloyd99}
\bibinfo{author}{\bibfnamefont{S.}~\bibnamefont{Lloyd}} \bibnamefont{and}
  \bibinfo{author}{\bibfnamefont{S.}~\bibnamefont{Braunstein}},
  \bibinfo{journal}{Phys.Rev. Lett.} \textbf{\bibinfo{volume}{82}},
  \bibinfo{pages}{1784} (\bibinfo{year}{1999}).

\bibitem[{\citenamefont{Raussendorf and Briegel}(2001)}]{raussendorf01}
\bibinfo{author}{\bibfnamefont{R.}~\bibnamefont{Raussendorf}} \bibnamefont{and}
  \bibinfo{author}{\bibfnamefont{H.~J.} \bibnamefont{Briegel}},
  \bibinfo{journal}{Phys. Rev. Lett.} \textbf{\bibinfo{volume}{86}},
  \bibinfo{pages}{5188} (\bibinfo{year}{2001}).

\bibitem[{\citenamefont{Menicucci et~al.}(2006)\citenamefont{Menicucci, van
  Loock, Gu, Weedbrook, Ralph, and Nielsen}}]{menicucci06}
\bibinfo{author}{\bibfnamefont{N.~C.} \bibnamefont{Menicucci}},
  \bibinfo{author}{\bibfnamefont{P.}~\bibnamefont{van Loock}},
  \bibinfo{author}{\bibfnamefont{M.}~\bibnamefont{Gu}},
  \bibinfo{author}{\bibfnamefont{C.}~\bibnamefont{Weedbrook}},
  \bibinfo{author}{\bibfnamefont{T.~C.} \bibnamefont{Ralph}}, \bibnamefont{and}
  \bibinfo{author}{\bibfnamefont{M.~A.} \bibnamefont{Nielsen}},
  \bibinfo{journal}{Phys. Rev. Lett.} \textbf{\bibinfo{volume}{97}},
  \bibinfo{pages}{110501} (\bibinfo{year}{2006}).

\bibitem[{\citenamefont{van Loock et~al.}(2007)\citenamefont{van Loock,
  Weedbrook, and Gu}}]{vanloock07}
\bibinfo{author}{\bibfnamefont{P.}~\bibnamefont{van Loock}},
  \bibinfo{author}{\bibfnamefont{C.}~\bibnamefont{Weedbrook}},
  \bibnamefont{and} \bibinfo{author}{\bibfnamefont{M.}~\bibnamefont{Gu}},
  \bibinfo{journal}{Phys. Rev. A} \textbf{\bibinfo{volume}{76}},
  \bibinfo{pages}{032321} (\bibinfo{year}{2007}).

\bibitem[{\citenamefont{Gu et~al.}(2009)\citenamefont{Gu, Weedbrook, Menicucci,
  Ralph, and van Loock}}]{gu09}
\bibinfo{author}{\bibfnamefont{M.}~\bibnamefont{Gu}},
  \bibinfo{author}{\bibfnamefont{C.}~\bibnamefont{Weedbrook}},
  \bibinfo{author}{\bibfnamefont{N.~C.} \bibnamefont{Menicucci}},
  \bibinfo{author}{\bibfnamefont{T.~C.} \bibnamefont{Ralph}}, \bibnamefont{and}
  \bibinfo{author}{\bibfnamefont{P.}~\bibnamefont{van Loock}},
  \bibinfo{journal}{Phys. Rev. A} \textbf{\bibinfo{volume}{79}},
  \bibinfo{pages}{062318} (\bibinfo{year}{2009}).

\bibitem[{\citenamefont{van Loock}(2010)}]{vanloock10}
\bibinfo{author}{\bibfnamefont{P.}~\bibnamefont{van Loock}}
  (\bibinfo{year}{2010}), \eprint{arXiv:1002.4788}.

\bibitem[{\citenamefont{Braunstein and van Loock}(2005)}]{braunstein05}
\bibinfo{author}{\bibfnamefont{S.~L.} \bibnamefont{Braunstein}}
  \bibnamefont{and} \bibinfo{author}{\bibfnamefont{P.}~\bibnamefont{van
  Loock}}, \bibinfo{journal}{Rev. Mod. Phys.} \textbf{\bibinfo{volume}{77}},
  \bibinfo{pages}{513} (\bibinfo{year}{2005}).

\bibitem[{\citenamefont{Vaziri et~al.}(2002)\citenamefont{Vaziri, Weihs, and
  Zeilinger}}]{vaziri02}
\bibinfo{author}{\bibfnamefont{A.}~\bibnamefont{Vaziri}},
  \bibinfo{author}{\bibfnamefont{G.}~\bibnamefont{Weihs}}, \bibnamefont{and}
  \bibinfo{author}{\bibfnamefont{A.}~\bibnamefont{Zeilinger}},
  \bibinfo{journal}{Phys. Rev. Lett.} \textbf{\bibinfo{volume}{89}},
  \bibinfo{pages}{240401} (\bibinfo{year}{2002}).

\bibitem[{\citenamefont{Vaziri et~al.}(2003)\citenamefont{Vaziri, Pan,
  Jennewein, Weihs, and Zeilinger}}]{vaziri03}
\bibinfo{author}{\bibfnamefont{A.}~\bibnamefont{Vaziri}},
  \bibinfo{author}{\bibfnamefont{J.-W.} \bibnamefont{Pan}},
  \bibinfo{author}{\bibfnamefont{T.}~\bibnamefont{Jennewein}},
  \bibinfo{author}{\bibfnamefont{G.}~\bibnamefont{Weihs}}, \bibnamefont{and}
  \bibinfo{author}{\bibfnamefont{A.}~\bibnamefont{Zeilinger}},
  \bibinfo{journal}{Phys. Rev. Lett.} \textbf{\bibinfo{volume}{91}},
  \bibinfo{pages}{227902} (\bibinfo{year}{2003}).

\bibitem[{\citenamefont{de~Oliveira et~al.}(2005)\citenamefont{de~Oliveira,
  Walborn, and Monken}}]{oliveira05}
\bibinfo{author}{\bibfnamefont{A.~N.} \bibnamefont{de~Oliveira}},
  \bibinfo{author}{\bibfnamefont{S.~P.} \bibnamefont{Walborn}},
  \bibnamefont{and} \bibinfo{author}{\bibfnamefont{C.~H.}
  \bibnamefont{Monken}}, \bibinfo{journal}{Journal of Optics B: Quantum and
  Semiclassical Optics} \textbf{\bibinfo{volume}{7}}, \bibinfo{pages}{288}
  (\bibinfo{year}{2005}).

\bibitem[{\citenamefont{Abouraddy et~al.}(2007)\citenamefont{Abouraddy,
  Yarnall, Saleh, and Teich}}]{abouraddy07}
\bibinfo{author}{\bibfnamefont{A.~F.} \bibnamefont{Abouraddy}},
  \bibinfo{author}{\bibfnamefont{T.}~\bibnamefont{Yarnall}},
  \bibinfo{author}{\bibfnamefont{B.~E.~A.} \bibnamefont{Saleh}},
  \bibnamefont{and} \bibinfo{author}{\bibfnamefont{M.~C.} \bibnamefont{Teich}},
  \bibinfo{journal}{Phys. Rev. A} \textbf{\bibinfo{volume}{75}},
  \bibinfo{eid}{052114} (pages~\bibinfo{numpages}{14}) (\bibinfo{year}{2007}).

\bibitem[{\citenamefont{Yarnall et~al.}(2007)\citenamefont{Yarnall, Abouraddy,
  Saleh, and Teich}}]{yarnall07}
\bibinfo{author}{\bibfnamefont{T.}~\bibnamefont{Yarnall}},
  \bibinfo{author}{\bibfnamefont{A.~F.} \bibnamefont{Abouraddy}},
  \bibinfo{author}{\bibfnamefont{B.~E.~A.} \bibnamefont{Saleh}},
  \bibnamefont{and} \bibinfo{author}{\bibfnamefont{M.~C.} \bibnamefont{Teich}},
  \bibinfo{journal}{Physical Review Letters} \textbf{\bibinfo{volume}{99}},
  \bibinfo{eid}{170408} (pages~\bibinfo{numpages}{4}) (\bibinfo{year}{2007}).

\bibitem[{\citenamefont{Howell et~al.}(2004)\citenamefont{Howell, Bennink,
  Bentley, and Boyd}}]{howell04}
\bibinfo{author}{\bibfnamefont{J.~C.} \bibnamefont{Howell}},
  \bibinfo{author}{\bibfnamefont{R.~S.} \bibnamefont{Bennink}},
  \bibinfo{author}{\bibfnamefont{S.~J.} \bibnamefont{Bentley}},
  \bibnamefont{and} \bibinfo{author}{\bibfnamefont{R.~W.} \bibnamefont{Boyd}},
  \bibinfo{journal}{Phys. Rev. Lett} \textbf{\bibinfo{volume}{92}},
  \bibinfo{pages}{210403} (\bibinfo{year}{2004}).

\bibitem[{\citenamefont{D'Angelo et~al.}(2004)\citenamefont{D'Angelo, Kim,
  Kulik, and Shih}}]{dangelo04}
\bibinfo{author}{\bibfnamefont{M.}~\bibnamefont{D'Angelo}},
  \bibinfo{author}{\bibfnamefont{Y.-H.} \bibnamefont{Kim}},
  \bibinfo{author}{\bibfnamefont{S.~P.} \bibnamefont{Kulik}}, \bibnamefont{and}
  \bibinfo{author}{\bibfnamefont{Y.}~\bibnamefont{Shih}},
  \bibinfo{journal}{Phys. Rev. Lett.} \textbf{\bibinfo{volume}{92}},
  \bibinfo{pages}{233601} (\bibinfo{year}{2004}).

\bibitem[{\citenamefont{Tasca et~al.}(2008)\citenamefont{Tasca, Walborn,
  Toscano, and Ribeiro}}]{tasca08}
\bibinfo{author}{\bibfnamefont{D.~S.} \bibnamefont{Tasca}},
  \bibinfo{author}{\bibfnamefont{S.~P.} \bibnamefont{Walborn}},
  \bibinfo{author}{\bibfnamefont{F.}~\bibnamefont{Toscano}}, \bibnamefont{and}
  \bibinfo{author}{\bibfnamefont{P.~H.~S.} \bibnamefont{Ribeiro}},
  \bibinfo{journal}{Phys. Rev. A} \textbf{\bibinfo{volume}{78}},
  \bibinfo{pages}{010304(R)} (\bibinfo{year}{2008}).

\bibitem[{\citenamefont{Tasca et~al.}(2009{\natexlab{a}})\citenamefont{Tasca,
  Walborn, Ribeiro, Toscano, and Pellat-Finet}}]{tasca09}
\bibinfo{author}{\bibfnamefont{D.~S.} \bibnamefont{Tasca}},
  \bibinfo{author}{\bibfnamefont{S.~P.} \bibnamefont{Walborn}},
  \bibinfo{author}{\bibfnamefont{P.~H.~S.} \bibnamefont{Ribeiro}},
  \bibinfo{author}{\bibfnamefont{F.}~\bibnamefont{Toscano}}, \bibnamefont{and}
  \bibinfo{author}{\bibfnamefont{P.}~\bibnamefont{Pellat-Finet}},
  \bibinfo{journal}{Phys. Rev. A} \textbf{\bibinfo{volume}{79}},
  \bibinfo{pages}{033801} (\bibinfo{year}{2009}{\natexlab{a}}).

\bibitem[{\citenamefont{Di~Lorenzo~Pires
  et~al.}(2009)\citenamefont{Di~Lorenzo~Pires, Monken, and van
  Exter}}]{pires09c}
\bibinfo{author}{\bibfnamefont{H.}~\bibnamefont{Di~Lorenzo~Pires}},
  \bibinfo{author}{\bibfnamefont{C.~H.} \bibnamefont{Monken}},
  \bibnamefont{and} \bibinfo{author}{\bibfnamefont{M.~P.} \bibnamefont{van
  Exter}}, \bibinfo{journal}{Phys. Rev. A} \textbf{\bibinfo{volume}{80}},
  \bibinfo{pages}{022307} (\bibinfo{year}{2009}).

\bibitem[{\citenamefont{Almeida et~al.}(2005)\citenamefont{Almeida, Walborn,
  and Ribeiro}}]{almeida05}
\bibinfo{author}{\bibfnamefont{M.~P.} \bibnamefont{Almeida}},
  \bibinfo{author}{\bibfnamefont{S.~P.} \bibnamefont{Walborn}},
  \bibnamefont{and} \bibinfo{author}{\bibfnamefont{P.~H.~S.}
  \bibnamefont{Ribeiro}}, \bibinfo{journal}{Phys. Rev. A}
  \textbf{\bibinfo{volume}{72}}, \bibinfo{pages}{022313}
  (\bibinfo{year}{2005}).

\bibitem[{\citenamefont{Walborn et~al.}(2006)\citenamefont{Walborn, Lemelle,
  Almeida, and Ribeiro}}]{walborn06}
\bibinfo{author}{\bibfnamefont{S.~P.} \bibnamefont{Walborn}},
  \bibinfo{author}{\bibfnamefont{D.~S.} \bibnamefont{Lemelle}},
  \bibinfo{author}{\bibfnamefont{M.~P.} \bibnamefont{Almeida}},
  \bibnamefont{and} \bibinfo{author}{\bibfnamefont{P.~H.~S.}
  \bibnamefont{Ribeiro}}, \bibinfo{journal}{Phys.Rev. Lett.}
  \textbf{\bibinfo{volume}{96}}, \bibinfo{pages}{090501}
  (\bibinfo{year}{2006}).

\bibitem[{\citenamefont{Walborn et~al.}(2008)\citenamefont{Walborn, Lemelle,
  Tasca, and Ribeiro}}]{walborn08}
\bibinfo{author}{\bibfnamefont{S.~P.} \bibnamefont{Walborn}},
  \bibinfo{author}{\bibfnamefont{D.~S.} \bibnamefont{Lemelle}},
  \bibinfo{author}{\bibfnamefont{D.~S.} \bibnamefont{Tasca}}, \bibnamefont{and}
  \bibinfo{author}{\bibfnamefont{P.~H.~S.} \bibnamefont{Ribeiro}},
  \bibinfo{journal}{Phys. Rev. A} \textbf{\bibinfo{volume}{77}},
  \bibinfo{pages}{062323} (\bibinfo{year}{2008}).

\bibitem[{\citenamefont{Tasca et~al.}(2009{\natexlab{b}})\citenamefont{Tasca,
  Walborn, Toscano, and Souto~Ribeiro}}]{tasca09b}
\bibinfo{author}{\bibfnamefont{D.~S.} \bibnamefont{Tasca}},
  \bibinfo{author}{\bibfnamefont{S.~P.} \bibnamefont{Walborn}},
  \bibinfo{author}{\bibfnamefont{F.}~\bibnamefont{Toscano}}, \bibnamefont{and}
  \bibinfo{author}{\bibfnamefont{P.~H.} \bibnamefont{Souto~Ribeiro}},
  \bibinfo{journal}{Phys. Rev. A} \textbf{\bibinfo{volume}{80}},
  \bibinfo{pages}{030101} (\bibinfo{year}{2009}{\natexlab{b}}).

\bibitem[{\citenamefont{Gomes et~al.}(2009{\natexlab{a}})\citenamefont{Gomes,
  Salles, Toscano, Ribeiro, and Walborn}}]{gomes09b}
\bibinfo{author}{\bibfnamefont{R.~M.} \bibnamefont{Gomes}},
  \bibinfo{author}{\bibfnamefont{A.}~\bibnamefont{Salles}},
  \bibinfo{author}{\bibfnamefont{F.}~\bibnamefont{Toscano}},
  \bibinfo{author}{\bibfnamefont{P.~H.~S.} \bibnamefont{Ribeiro}},
  \bibnamefont{and} \bibinfo{author}{\bibfnamefont{S.~P.}
  \bibnamefont{Walborn}}, \bibinfo{journal}{Proc. Nat. Acad. Sci.}
  \textbf{\bibinfo{volume}{106}}, \bibinfo{pages}{21517}
  (\bibinfo{year}{2009}{\natexlab{a}}).

\bibitem[{\citenamefont{Walborn
  et~al.}(2003{\natexlab{a}})\citenamefont{Walborn, de~Oliveira, P\'adua, and
  Monken}}]{walborn03a}
\bibinfo{author}{\bibfnamefont{S.~P.} \bibnamefont{Walborn}},
  \bibinfo{author}{\bibfnamefont{A.~N.} \bibnamefont{de~Oliveira}},
  \bibinfo{author}{\bibfnamefont{S.}~\bibnamefont{P\'adua}}, \bibnamefont{and}
  \bibinfo{author}{\bibfnamefont{C.~H.} \bibnamefont{Monken}},
  \bibinfo{journal}{Phys. Rev. Lett} \textbf{\bibinfo{volume}{90}},
  \bibinfo{pages}{143601} (\bibinfo{year}{2003}{\natexlab{a}}).

\bibitem[{\citenamefont{Walborn
  et~al.}(2003{\natexlab{b}})\citenamefont{Walborn, de~Oliveira, P\'adua, and
  Monken}}]{walborn03b}
\bibinfo{author}{\bibfnamefont{S.~P.} \bibnamefont{Walborn}},
  \bibinfo{author}{\bibfnamefont{A.~N.} \bibnamefont{de~Oliveira}},
  \bibinfo{author}{\bibfnamefont{S.}~\bibnamefont{P\'adua}}, \bibnamefont{and}
  \bibinfo{author}{\bibfnamefont{C.~H.} \bibnamefont{Monken}},
  \bibinfo{journal}{Europhys. Lett} \textbf{\bibinfo{volume}{62}},
  \bibinfo{pages}{161} (\bibinfo{year}{2003}{\natexlab{b}}).

\bibitem[{\citenamefont{Caetano and Souto~Ribeiro}(2003)}]{caetano03}
\bibinfo{author}{\bibfnamefont{D.~P.} \bibnamefont{Caetano}} \bibnamefont{and}
  \bibinfo{author}{\bibfnamefont{P.~H.} \bibnamefont{Souto~Ribeiro}},
  \bibinfo{journal}{Phys. Rev. A} \textbf{\bibinfo{volume}{68}},
  \bibinfo{pages}{043806} (\bibinfo{year}{2003}).

\bibitem[{\citenamefont{Nogueira et~al.}(2004)\citenamefont{Nogueira, Walborn,
  P\'adua, and Monken}}]{nogueira04a}
\bibinfo{author}{\bibfnamefont{W.~A.~T.} \bibnamefont{Nogueira}},
  \bibinfo{author}{\bibfnamefont{S.~P.} \bibnamefont{Walborn}},
  \bibinfo{author}{\bibfnamefont{S.}~\bibnamefont{P\'adua}}, \bibnamefont{and}
  \bibinfo{author}{\bibfnamefont{C.~H.} \bibnamefont{Monken}},
  \bibinfo{journal}{Phys. Rev. Lett.} \textbf{\bibinfo{volume}{92}},
  \bibinfo{pages}{043602} (\bibinfo{year}{2004}).

\bibitem[{\citenamefont{Gomes et~al.}(2009{\natexlab{b}})\citenamefont{Gomes,
  Salles, Toscano, Ribeiro, and Walborn}}]{gomes09a}
\bibinfo{author}{\bibfnamefont{R.~M.} \bibnamefont{Gomes}},
  \bibinfo{author}{\bibfnamefont{A.}~\bibnamefont{Salles}},
  \bibinfo{author}{\bibfnamefont{F.}~\bibnamefont{Toscano}},
  \bibinfo{author}{\bibfnamefont{P.~H.~S.} \bibnamefont{Ribeiro}},
  \bibnamefont{and} \bibinfo{author}{\bibfnamefont{S.~P.}
  \bibnamefont{Walborn}}, \bibinfo{journal}{Phys. Rev. Lett.}
  \textbf{\bibinfo{volume}{103}}, \bibinfo{pages}{033602}
  (\bibinfo{year}{2009}{\natexlab{b}}).

\bibitem[{\citenamefont{Mair et~al.}(2001)\citenamefont{Mair, Vaziri, Weihs,
  and Zeilinger}}]{mair01}
\bibinfo{author}{\bibfnamefont{A.}~\bibnamefont{Mair}},
  \bibinfo{author}{\bibfnamefont{A.}~\bibnamefont{Vaziri}},
  \bibinfo{author}{\bibfnamefont{G.}~\bibnamefont{Weihs}}, \bibnamefont{and}
  \bibinfo{author}{\bibfnamefont{A.}~\bibnamefont{Zeilinger}},
  \bibinfo{journal}{Nature} \textbf{\bibinfo{volume}{412}},
  \bibinfo{pages}{313} (\bibinfo{year}{2001}).

\bibitem[{\citenamefont{Walborn et~al.}(2004)\citenamefont{Walborn,
  de~Oliveira, Thebaldi, and Monken}}]{walborn04a}
\bibinfo{author}{\bibfnamefont{S.~P.} \bibnamefont{Walborn}},
  \bibinfo{author}{\bibfnamefont{A.~N.} \bibnamefont{de~Oliveira}},
  \bibinfo{author}{\bibfnamefont{R.~S.} \bibnamefont{Thebaldi}},
  \bibnamefont{and} \bibinfo{author}{\bibfnamefont{C.~H.}
  \bibnamefont{Monken}}, \bibinfo{journal}{Phys. Rev. A}
  \textbf{\bibinfo{volume}{69}}, \bibinfo{pages}{023811}
  (\bibinfo{year}{2004}).

\bibitem[{\citenamefont{Peeters et~al.}(2007)\citenamefont{Peeters, Verstegen,
  and van Exter}}]{peeters07}
\bibinfo{author}{\bibfnamefont{W.~H.} \bibnamefont{Peeters}},
  \bibinfo{author}{\bibfnamefont{E.~J.~K.} \bibnamefont{Verstegen}},
  \bibnamefont{and} \bibinfo{author}{\bibfnamefont{M.~P.} \bibnamefont{van
  Exter}}, \bibinfo{journal}{Phys. Rev. A} \textbf{\bibinfo{volume}{76}},
  \bibinfo{pages}{042302} (\bibinfo{year}{2007}).

\bibitem[{\citenamefont{Di~Lorenzo~Pires
  et~al.}(2010)\citenamefont{Di~Lorenzo~Pires, Florijn, and van
  Exter}}]{pires10}
\bibinfo{author}{\bibfnamefont{H.}~\bibnamefont{Di~Lorenzo~Pires}},
  \bibinfo{author}{\bibfnamefont{H.~C.~B.} \bibnamefont{Florijn}},
  \bibnamefont{and} \bibinfo{author}{\bibfnamefont{M.~P.} \bibnamefont{van
  Exter}}, \bibinfo{journal}{Phys. Rev. Lett.} \textbf{\bibinfo{volume}{104}},
  \bibinfo{pages}{020505} (\bibinfo{year}{2010}).

\bibitem[{\citenamefont{Leach et~al.}(2002)\citenamefont{Leach, Padgett,
  Barnett, Franke-Arnold, and Courtial}}]{leach02}
\bibinfo{author}{\bibfnamefont{J.}~\bibnamefont{Leach}},
  \bibinfo{author}{\bibfnamefont{M.~J.} \bibnamefont{Padgett}},
  \bibinfo{author}{\bibfnamefont{S.~M.} \bibnamefont{Barnett}},
  \bibinfo{author}{\bibfnamefont{S.}~\bibnamefont{Franke-Arnold}},
  \bibnamefont{and} \bibinfo{author}{\bibfnamefont{J.}~\bibnamefont{Courtial}},
  \bibinfo{journal}{Phys. Rev. Lett.} \textbf{\bibinfo{volume}{88}},
  \bibinfo{pages}{257901} (\bibinfo{year}{2002}).

\bibitem[{\citenamefont{Wei et~al.}(2003)\citenamefont{Wei, Xue, Leach,
  Padgett, Barnett, Franke-Arnold, Yao, and Courtial}}]{wei03}
\bibinfo{author}{\bibfnamefont{H.}~\bibnamefont{Wei}},
  \bibinfo{author}{\bibfnamefont{X.}~\bibnamefont{Xue}},
  \bibinfo{author}{\bibfnamefont{J.}~\bibnamefont{Leach}},
  \bibinfo{author}{\bibfnamefont{M.~J.} \bibnamefont{Padgett}},
  \bibinfo{author}{\bibfnamefont{S.~M.} \bibnamefont{Barnett}},
  \bibinfo{author}{\bibfnamefont{S.}~\bibnamefont{Franke-Arnold}},
  \bibinfo{author}{\bibfnamefont{E.}~\bibnamefont{Yao}}, \bibnamefont{and}
  \bibinfo{author}{\bibfnamefont{J.}~\bibnamefont{Courtial}},
  \bibinfo{journal}{Optics Comm.} \textbf{\bibinfo{volume}{223}},
  \bibinfo{pages}{117} (\bibinfo{year}{2003}).

\bibitem[{\citenamefont{Berkhout et~al.}(2010)\citenamefont{Berkhout, Lavery,
  Courtial, Beijersbergen, and Padgett}}]{berkhout10}
\bibinfo{author}{\bibfnamefont{G.~C.~G.} \bibnamefont{Berkhout}},
  \bibinfo{author}{\bibfnamefont{M.~P.~J.} \bibnamefont{Lavery}},
  \bibinfo{author}{\bibfnamefont{J.}~\bibnamefont{Courtial}},
  \bibinfo{author}{\bibfnamefont{M.~W.} \bibnamefont{Beijersbergen}},
  \bibnamefont{and} \bibinfo{author}{\bibfnamefont{M.~J.}
  \bibnamefont{Padgett}}, \bibinfo{journal}{Phys. Rev. Lett.}
  \textbf{\bibinfo{volume}{105}}, \bibinfo{pages}{153601}
  (\bibinfo{year}{2010}).

\bibitem[{\citenamefont{Oemrawsingh et~al.}(2005)\citenamefont{Oemrawsingh,
  X.~Ma, Aiello, Eliel, 't~Hooft, and Woerdman}}]{oemrawsingh05}
\bibinfo{author}{\bibfnamefont{S.~S.~R.} \bibnamefont{Oemrawsingh}},
  \bibinfo{author}{\bibfnamefont{D.~V.} \bibnamefont{X.~Ma}},
  \bibinfo{author}{\bibfnamefont{A.}~\bibnamefont{Aiello}},
  \bibinfo{author}{\bibfnamefont{E.~R.} \bibnamefont{Eliel}},
  \bibinfo{author}{\bibfnamefont{G.~W.} \bibnamefont{'t~Hooft}},
  \bibnamefont{and} \bibinfo{author}{\bibfnamefont{J.~P.}
  \bibnamefont{Woerdman}}, \bibinfo{journal}{Phys. Rev. Lett.}
  \textbf{\bibinfo{volume}{95}}, \bibinfo{pages}{240501}
  (\bibinfo{year}{2005}).

\bibitem[{\citenamefont{Pors et~al.}(2008{\natexlab{a}})\citenamefont{Pors,
  Aiello, Oemrawsingh, van Exter, Eliel, and Woerdman}}]{pors08}
\bibinfo{author}{\bibfnamefont{J.~B.} \bibnamefont{Pors}},
  \bibinfo{author}{\bibfnamefont{A.}~\bibnamefont{Aiello}},
  \bibinfo{author}{\bibfnamefont{S.~S.~R.} \bibnamefont{Oemrawsingh}},
  \bibinfo{author}{\bibfnamefont{M.~P.} \bibnamefont{van Exter}},
  \bibinfo{author}{\bibfnamefont{E.~R.} \bibnamefont{Eliel}}, \bibnamefont{and}
  \bibinfo{author}{\bibfnamefont{J.~P.} \bibnamefont{Woerdman}},
  \bibinfo{journal}{Phys. Rev. A} \textbf{\bibinfo{volume}{77}},
  \bibinfo{eid}{033845} (pages~\bibinfo{numpages}{6})
  (\bibinfo{year}{2008}{\natexlab{a}}).

\bibitem[{\citenamefont{Pors et~al.}(2008{\natexlab{b}})\citenamefont{Pors,
  Oemrawsingh, Aiello, van Exter, Eliel, 't~Hooft, and Woerdman}}]{pors08b}
\bibinfo{author}{\bibfnamefont{J.~B.} \bibnamefont{Pors}},
  \bibinfo{author}{\bibfnamefont{S.~S.~R.} \bibnamefont{Oemrawsingh}},
  \bibinfo{author}{\bibfnamefont{A.}~\bibnamefont{Aiello}},
  \bibinfo{author}{\bibfnamefont{M.~P.} \bibnamefont{van Exter}},
  \bibinfo{author}{\bibfnamefont{E.~R.} \bibnamefont{Eliel}},
  \bibinfo{author}{\bibfnamefont{G.~W.} \bibnamefont{'t~Hooft}},
  \bibnamefont{and} \bibinfo{author}{\bibfnamefont{J.~P.}
  \bibnamefont{Woerdman}}, \bibinfo{journal}{Physical Review Letters}
  \textbf{\bibinfo{volume}{101}}, \bibinfo{eid}{120502}
  (\bibinfo{year}{2008}{\natexlab{b}}).

\bibitem[{\citenamefont{Knill et~al.}(2001)\citenamefont{Knill, Laflamme, and
  Milburn}}]{klm01}
\bibinfo{author}{\bibfnamefont{E.}~\bibnamefont{Knill}},
  \bibinfo{author}{\bibfnamefont{R.}~\bibnamefont{Laflamme}}, \bibnamefont{and}
  \bibinfo{author}{\bibfnamefont{G.~J.} \bibnamefont{Milburn}},
  \bibinfo{journal}{Nature (London)} \textbf{\bibinfo{volume}{409}},
  \bibinfo{pages}{46} (\bibinfo{year}{2001}).

\bibitem[{\citenamefont{Gottesman et~al.}(2001)\citenamefont{Gottesman, Kitaev,
  and Preskill}}]{gkp01}
\bibinfo{author}{\bibfnamefont{D.}~\bibnamefont{Gottesman}},
  \bibinfo{author}{\bibfnamefont{A.}~\bibnamefont{Kitaev}}, \bibnamefont{and}
  \bibinfo{author}{\bibfnamefont{J.}~\bibnamefont{Preskill}},
  \bibinfo{journal}{Phys. Rev. A} \textbf{\bibinfo{volume}{64}},
  \bibinfo{pages}{012310} (\bibinfo{year}{2001}).

\bibitem[{\citenamefont{Adcock et~al.}(2009)\citenamefont{Adcock, Hoyer, and
  Sanders}}]{adcock09}
\bibinfo{author}{\bibfnamefont{M.~R.~A.} \bibnamefont{Adcock}},
  \bibinfo{author}{\bibfnamefont{P.}~\bibnamefont{Hoyer}}, \bibnamefont{and}
  \bibinfo{author}{\bibfnamefont{B.~C.} \bibnamefont{Sanders}},
  \bibinfo{journal}{New J. Phys.} \textbf{\bibinfo{volume}{11}},
  \bibinfo{pages}{103035} (\bibinfo{year}{2009}).

\bibitem[{\citenamefont{Ohliger et~al.}(2010)\citenamefont{Ohliger, Kieling,
  and Eisert}}]{eisert10}
\bibinfo{author}{\bibfnamefont{M.}~\bibnamefont{Ohliger}},
  \bibinfo{author}{\bibfnamefont{K.}~\bibnamefont{Kieling}}, \bibnamefont{and}
  \bibinfo{author}{\bibfnamefont{J.}~\bibnamefont{Eisert}},
  \bibinfo{journal}{Phys. Rev. A} \textbf{\bibinfo{volume}{82}},
  \bibinfo{pages}{042336} (\bibinfo{year}{2010}).

\bibitem[{\citenamefont{Cable and Browne}(2010)}]{cable10}
\bibinfo{author}{\bibfnamefont{H.}~\bibnamefont{Cable}} \bibnamefont{and}
  \bibinfo{author}{\bibfnamefont{D.~E.} \bibnamefont{Browne}},
  \bibinfo{journal}{New J. Phys.} \textbf{\bibinfo{volume}{12}},
  \bibinfo{pages}{113046} (\bibinfo{year}{2010}).

\bibitem[{\citenamefont{Popp et~al.}(2005)\citenamefont{Popp, Verstraete,
  Mart\'\i{}n-Delgado, and Cirac}}]{verstraete05}
\bibinfo{author}{\bibfnamefont{M.}~\bibnamefont{Popp}},
  \bibinfo{author}{\bibfnamefont{F.}~\bibnamefont{Verstraete}},
  \bibinfo{author}{\bibfnamefont{M.~A.} \bibnamefont{Mart\'\i{}n-Delgado}},
  \bibnamefont{and} \bibinfo{author}{\bibfnamefont{J.~I.} \bibnamefont{Cirac}},
  \bibinfo{journal}{Phys. Rev. A} \textbf{\bibinfo{volume}{71}},
  \bibinfo{pages}{042306} (\bibinfo{year}{2005}).

\bibitem[{\citenamefont{Lvovsky and Raymer}(2009)}]{lvovsky09}
\bibinfo{author}{\bibfnamefont{A.~I.} \bibnamefont{Lvovsky}} \bibnamefont{and}
  \bibinfo{author}{\bibfnamefont{M.~G.} \bibnamefont{Raymer}},
  \bibinfo{journal}{Rev. Mod. Phys.} \textbf{\bibinfo{volume}{81}},
  \bibinfo{pages}{299} (\bibinfo{year}{2009}).

\bibitem[{\citenamefont{Marcuse}(1982)}]{marcuse}
\bibinfo{author}{\bibfnamefont{D.}~\bibnamefont{Marcuse}},
  \emph{\bibinfo{title}{Light Transmission Optics}} (\bibinfo{publisher}{Van
  Nostrand Reinhold Company}, \bibinfo{address}{New York},
  \bibinfo{year}{1982}).

\bibitem[{\citenamefont{Stoler}(1981)}]{stoler81}
\bibinfo{author}{\bibfnamefont{D.}~\bibnamefont{Stoler}}, \bibinfo{journal}{J.
  Opt. Soc. Am.} \textbf{\bibinfo{volume}{71}}, \bibinfo{pages}{334}
  (\bibinfo{year}{1981}).

\bibitem[{\citenamefont{Guillemin and Sternberg}(1984)}]{guillemin}
\bibinfo{author}{\bibfnamefont{V.}~\bibnamefont{Guillemin}} \bibnamefont{and}
  \bibinfo{author}{\bibfnamefont{S.}~\bibnamefont{Sternberg}},
  \emph{\bibinfo{title}{Symplectic techniques in physics}}
  (\bibinfo{publisher}{Cambridge University Press},
  \bibinfo{address}{Cambridge}, \bibinfo{year}{1984}).

\bibitem[{\citenamefont{Lohmann}(1995)}]{lohmann93}
\bibinfo{author}{\bibfnamefont{A.~W.} \bibnamefont{Lohmann}},
  \bibinfo{journal}{Optics Comm.} \textbf{\bibinfo{volume}{115}},
  \bibinfo{pages}{437} (\bibinfo{year}{1995}).

\bibitem[{\citenamefont{Pellat-Finet}(1994)}]{pellat-finet94}
\bibinfo{author}{\bibfnamefont{P.}~\bibnamefont{Pellat-Finet}},
  \bibinfo{journal}{Opt. Lett.} \textbf{\bibinfo{volume}{19}},
  \bibinfo{pages}{1388} (\bibinfo{year}{1994}).

\bibitem[{\citenamefont{Ozaktas et~al.}(2001)\citenamefont{Ozaktas, Zalevsky,
  and Kutay}}]{ozaktas01}
\bibinfo{author}{\bibfnamefont{H.~M.} \bibnamefont{Ozaktas}},
  \bibinfo{author}{\bibfnamefont{Z.}~\bibnamefont{Zalevsky}}, \bibnamefont{and}
  \bibinfo{author}{\bibfnamefont{M.~A.} \bibnamefont{Kutay}},
  \emph{\bibinfo{title}{The Fractional Fourier Transform: with Applications in
  Optics and Signal Processing}} (\bibinfo{publisher}{John Wiley and Sons
  Ltd.}, \bibinfo{address}{West Sussex}, \bibinfo{year}{2001}).

\bibitem[{\citenamefont{Goodman}(1996)}]{goodman96}
\bibinfo{author}{\bibfnamefont{J.~W.} \bibnamefont{Goodman}},
  \emph{\bibinfo{title}{Introduction to Fourier Optics}}
  (\bibinfo{publisher}{Mc Graw Hill}, \bibinfo{address}{Boston},
  \bibinfo{year}{1996}).

\bibitem[{\citenamefont{Milburn}(1989)}]{milburn89}
\bibinfo{author}{\bibfnamefont{G.~J.} \bibnamefont{Milburn}},
  \bibinfo{journal}{Phys. Rev. Lett.} \textbf{\bibinfo{volume}{62}},
  \bibinfo{pages}{2124} (\bibinfo{year}{1989}).

\bibitem[{\citenamefont{O'Brien et~al.}(2003)\citenamefont{O'Brien, Pryde,
  White, Ralph, and Branning}}]{obrien03}
\bibinfo{author}{\bibfnamefont{J.~L.} \bibnamefont{O'Brien}},
  \bibinfo{author}{\bibfnamefont{G.~J.} \bibnamefont{Pryde}},
  \bibinfo{author}{\bibfnamefont{A.~G.} \bibnamefont{White}},
  \bibinfo{author}{\bibfnamefont{T.~C.} \bibnamefont{Ralph}}, \bibnamefont{and}
  \bibinfo{author}{\bibfnamefont{D.}~\bibnamefont{Branning}},
  \bibinfo{journal}{Nature} \textbf{\bibinfo{volume}{426}},
  \bibinfo{pages}{264} (\bibinfo{year}{2003}).

\bibitem[{\citenamefont{Langford et~al.}(2005)\citenamefont{Langford, Weinhold,
  Prevedel, Resch, Gilchrist, O\char39{}Brien, Pryde, and White}}]{langford05}
\bibinfo{author}{\bibfnamefont{N.~K.} \bibnamefont{Langford}},
  \bibinfo{author}{\bibfnamefont{T.~J.} \bibnamefont{Weinhold}},
  \bibinfo{author}{\bibfnamefont{R.}~\bibnamefont{Prevedel}},
  \bibinfo{author}{\bibfnamefont{K.~J.} \bibnamefont{Resch}},
  \bibinfo{author}{\bibfnamefont{A.}~\bibnamefont{Gilchrist}},
  \bibinfo{author}{\bibfnamefont{J.~L.} \bibnamefont{O\char39{}Brien}},
  \bibinfo{author}{\bibfnamefont{G.~J.} \bibnamefont{Pryde}}, \bibnamefont{and}
  \bibinfo{author}{\bibfnamefont{A.~G.} \bibnamefont{White}},
  \bibinfo{journal}{Phys. Rev. Lett.} \textbf{\bibinfo{volume}{95}},
  \bibinfo{pages}{210504} (\bibinfo{year}{2005}).

\bibitem[{\citenamefont{Kiesel et~al.}(2005)\citenamefont{Kiesel, Schmid,
  Weber, Ursin, and Weinfurter}}]{kiesel05}
\bibinfo{author}{\bibfnamefont{N.}~\bibnamefont{Kiesel}},
  \bibinfo{author}{\bibfnamefont{C.}~\bibnamefont{Schmid}},
  \bibinfo{author}{\bibfnamefont{U.}~\bibnamefont{Weber}},
  \bibinfo{author}{\bibfnamefont{R.}~\bibnamefont{Ursin}}, \bibnamefont{and}
  \bibinfo{author}{\bibfnamefont{H.}~\bibnamefont{Weinfurter}},
  \bibinfo{journal}{Phys. Rev. Lett.} \textbf{\bibinfo{volume}{95}},
  \bibinfo{pages}{210505} (\bibinfo{year}{2005}).

\bibitem[{\citenamefont{Okamoto et~al.}(2005)\citenamefont{Okamoto, Hofmann,
  Takeuchi, and Sasaki}}]{okamoto05}
\bibinfo{author}{\bibfnamefont{R.}~\bibnamefont{Okamoto}},
  \bibinfo{author}{\bibfnamefont{H.~F.} \bibnamefont{Hofmann}},
  \bibinfo{author}{\bibfnamefont{S.}~\bibnamefont{Takeuchi}}, \bibnamefont{and}
  \bibinfo{author}{\bibfnamefont{K.}~\bibnamefont{Sasaki}},
  \bibinfo{journal}{Phys. Rev. Lett.} \textbf{\bibinfo{volume}{95}},
  \bibinfo{pages}{210506} (\bibinfo{year}{2005}).

\bibitem[{\citenamefont{Lanyon et~al.}(2008)\citenamefont{Lanyon, Barbieri,
  Almeida, Jennewein, Ralph, Resch, Pryde, OÕBrien, Gilchrist, and
  White}}]{lanyon08}
\bibinfo{author}{\bibfnamefont{B.~P.} \bibnamefont{Lanyon}},
  \bibinfo{author}{\bibfnamefont{M.}~\bibnamefont{Barbieri}},
  \bibinfo{author}{\bibfnamefont{M.~P.} \bibnamefont{Almeida}},
  \bibinfo{author}{\bibfnamefont{T.}~\bibnamefont{Jennewein}},
  \bibinfo{author}{\bibfnamefont{T.~C.} \bibnamefont{Ralph}},
  \bibinfo{author}{\bibfnamefont{K.~J.} \bibnamefont{Resch}},
  \bibinfo{author}{\bibfnamefont{G.~J.} \bibnamefont{Pryde}},
  \bibinfo{author}{\bibfnamefont{J.~L.} \bibnamefont{OÕBrien}},
  \bibinfo{author}{\bibfnamefont{A.}~\bibnamefont{Gilchrist}},
  \bibnamefont{and} \bibinfo{author}{\bibfnamefont{A.~G.} \bibnamefont{White}},
  \bibinfo{journal}{Nature Physics} \textbf{\bibinfo{volume}{5}},
  \bibinfo{pages}{134} (\bibinfo{year}{2008}).

\bibitem[{\citenamefont{Mandel and Wolf}(1995)}]{mandel95}
\bibinfo{author}{\bibfnamefont{L.}~\bibnamefont{Mandel}} \bibnamefont{and}
  \bibinfo{author}{\bibfnamefont{E.}~\bibnamefont{Wolf}},
  \emph{\bibinfo{title}{Optical Coherence and Quantum Optics}}
  (\bibinfo{publisher}{Cambridge University Press}, \bibinfo{address}{New
  York}, \bibinfo{year}{1995}).

\bibitem[{\citenamefont{Walborn et~al.}(2010)\citenamefont{Walborn, Monken,
  P\'adua, and Ribeiro}}]{walborn10a}
\bibinfo{author}{\bibfnamefont{S.~P.} \bibnamefont{Walborn}},
  \bibinfo{author}{\bibfnamefont{C.~H.} \bibnamefont{Monken}},
  \bibinfo{author}{\bibfnamefont{S.}~\bibnamefont{P\'adua}}, \bibnamefont{and}
  \bibinfo{author}{\bibfnamefont{P.~H.~S.} \bibnamefont{Ribeiro}},
  \bibinfo{journal}{Phys. Rep.} \textbf{\bibinfo{volume}{495}},
  \bibinfo{pages}{87} (\bibinfo{year}{2010}).

\bibitem[{\citenamefont{Ceccarelli et~al.}(2009)\citenamefont{Ceccarelli,
  Vallone, De~Martini, Mataloni, and Cabello}}]{ceccarelli09}
\bibinfo{author}{\bibfnamefont{R.}~\bibnamefont{Ceccarelli}},
  \bibinfo{author}{\bibfnamefont{G.}~\bibnamefont{Vallone}},
  \bibinfo{author}{\bibfnamefont{F.}~\bibnamefont{De~Martini}},
  \bibinfo{author}{\bibfnamefont{P.}~\bibnamefont{Mataloni}}, \bibnamefont{and}
  \bibinfo{author}{\bibfnamefont{A.}~\bibnamefont{Cabello}},
  \bibinfo{journal}{Phys. Rev. Lett.} \textbf{\bibinfo{volume}{103}},
  \bibinfo{pages}{160401} (\bibinfo{year}{2009}).

\bibitem[{\citenamefont{Torres et~al.}(2003)\citenamefont{Torres, Deyanova,
  Torner, and Molina-Terriza}}]{torres03b}
\bibinfo{author}{\bibfnamefont{J.~P.} \bibnamefont{Torres}},
  \bibinfo{author}{\bibfnamefont{Y.}~\bibnamefont{Deyanova}},
  \bibinfo{author}{\bibfnamefont{L.}~\bibnamefont{Torner}}, \bibnamefont{and}
  \bibinfo{author}{\bibfnamefont{G.}~\bibnamefont{Molina-Terriza}},
  \bibinfo{journal}{Phys. Rev. A} \textbf{\bibinfo{volume}{67}},
  \bibinfo{pages}{052313} (\bibinfo{year}{2003}).

\bibitem[{\citenamefont{Monken et~al.}(1998)\citenamefont{Monken, Ribeiro, and
  P\'adua}}]{monken98a}
\bibinfo{author}{\bibfnamefont{C.~H.} \bibnamefont{Monken}},
  \bibinfo{author}{\bibfnamefont{P.~S.} \bibnamefont{Ribeiro}},
  \bibnamefont{and} \bibinfo{author}{\bibfnamefont{S.}~\bibnamefont{P\'adua}},
  \bibinfo{journal}{Phys. Rev. A.} \textbf{\bibinfo{volume}{57}},
  \bibinfo{pages}{3123} (\bibinfo{year}{1998}).

\end{thebibliography}


\end{document}